\documentclass[12pt,preprint]{aastex}
\usepackage{amsmath,amssymb,amsfonts,textcomp,color}
\usepackage{graphicx}
\usepackage{longtable}
\usepackage{subfigure}
\usepackage{ccaption}

\begin{document}
\title{Active Galactic Nuclei. VII. The Spacecraft Wobble and X{}-Ray Variability of Seyfert Galaxy NGC 5548}
\author{Ludwik Liszka}
\affil{Swedish Institute of Space Physics, SLU, 90183 Umea, Sweden}
\email{ludwik@irf.se}

\author{A. G. Pacholczyk}
\affil{Steward Observatory, The University of Arizona, Tucson, AZ 85721}
\email{agp@as.arizona.edur}
\and
\author{William R. Stoeger, S.J.}
\affil{Vatican Observatory Research Group, Steward Observatory, The University of Arizona, Tucson, AZ 85721}
\email{wstoeger@as.arizona.edu}

\begin{abstract}
In discussing the short{}-time variability of extragalactic X{}-ray sources, we focus on explaining how wavelet transform in conjunction with non{}-linear filtering methods are used to remove Poisson statistics and wobble{}-related variability from the data. This enables the resolution of their intrinsic stochastic, quasi{}-periodic and deterministic variable components, as well the determination of their persistence. As detailed examples we review and extend the application of these techniques to ROSAT data for the Seyfert I galaxy NGC 5548 and for the QSO 3C273. Besides illustrating the discriminating power of these methods, these treatments confirm the intrinsic character of the transient quasi{}-periodic and deterministic events in NGC 5548 and 3C273 and enable us to compare both the elementary events and the short{}-time variability in the two types of sources.
These methods can be applied in the investigation of intrinsic source variability in other spectral regions.
\end{abstract}
\keywords{galaxies, active {}-{}- galaxies, Seyfert {}-{}- methods, statistical {}-{}-QSO's {}-{}- X{}-rays, galaxies}

\section{Introduction}
{
A common belief in X{}-ray astronomy is that the spacecraft wobble precludes the possibility of obtaining correct information about the variability of observed sources.
\par}
{
The spacecraft wobble was introduced in order to prevent a given source from always being detected by the same pixel/pixels of the sensor. The linear wobble in ROSAT has been replaced by 2{}-dimensional ``dithering'' in the recent Chandra satellite, but the purpose of the technique remains unchanged. In ROSAT a consequence of the wobble is that the mesh wires in the entrance window of the instrument occult the observed source, thus introducing an instrumental variability. Such instrumental variability complicates observations, but it is far from destroying the information about the real variability of the source. On the contrary, in different physical measurements the emission from the source is chopped in order to increase the possibility of detection. In nature, certain species, like jumping spiders (\textit{Salticidae}) acquire visual data by sweeping an essentially linear retina back and forth perpendicularly to its larger dimension \citep{land69}. It has been demonstrated that the oscillating retina not only increases the resolution of the image, but also improves the over{}-all performance of vision, including its temporal resolution. This vision{}-enhancing principle is planned for use in the next generation of Mars rovers (New Scientist, 2001).
\par}

{
Modern signal processing techniques offer a wide range of possibilities for separating different components in the signal. In the case of X{}-ray observations the recorded signal is, as a rule, polluted by Poisson statistics, due to small numbers of recorded photons. However, during recent years a wavelet technique has been developed \citep{liszka99} which removes much of the Poisson statistics from the data. In the present work the question whether the spacecraft wobble obscures the intrinsic variability of the source will be addressed using recent information processing techniques. Data from two intense X{}-ray sources: Crab and NGC 5548 were used in the present study.
\par}
{
Variability is an important characteristic of many astronomical systems, and rapid, apparently random variability is distinctive of many galactic and extra{}-galactic objects which are thought to harbor black holes, or other types of compact objects \ {}-{}- X{}-ray binaries, QSO's, AGN and sources of gamma{}-ray bursts.It has proved very difficult to analyze in any reliable way the various components of the variability data from these objects in order to learn more about their underlying causes within the sources themselves. Here we review and extend the application we have made elsewhere \citep{liszka99,liszka00a,liszka00b} of wavelet transform along with non{}-linear filtering techniques to the rapid X{}-ray variability of Seyfert 1 (S1) galaxies and QSO's. In particular, we show how we have employed these methods \ to remove the Poisson{}-statistical and wobble{}-related variability from the ROSAT data for the S1 NGC 5548 and for the QSO 3C273. In doing so have confirmed the presence of intrinsic transient low{}-level deterministic and even quasi{}-periodic structures in data, which strongly indicate the occurrence of quasi{}-regular elementary events in the sources themselves.
\par}
{
These results were first reported in \citet{liszka00a}. This paper presents a more detailed, careful and improved treatment of the methods used and the results obtained, particularly of the wobble{}-related and Poisson{}-statistical influences which must be removed to facilitate the determination of the properties of these sources. Our work confirms and strengthens observationally based support for the intrinsic character of the quasi{}-periodic and deterministic events we first reported in \citet{liszka00b}, and significantly extends it to determination of the persistence of such events, their time{}-scales, and amplitudes. The methods we describe and apply here can obviously be applied to a wide{}-range of variable astronomical phenomena. Further confirmation of these results, of course, is crucial {}-{}- particularly by using data from more recent X{}-ray satellites, such as Chandra.
\par}
\begin{figure}[tb]
\center
\includegraphics[height=.57\hsize]{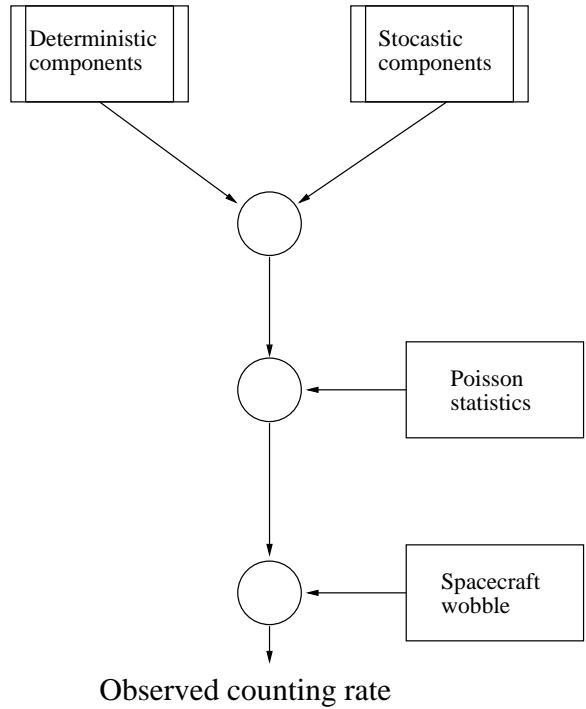}
\figcaption{Formation of the observed photon train\label{fig1}}
\end{figure}
{
The variability spectrum of an X{}-ray source may be a combination of deterministic and stochastic components (see the upper part of Fig.~\ref{fig1}). The combined photon flux is increasingly affected by the Poisson statistics with the increased distance from the source. Finally, the photon train, measured at the spacecraft, is modulated by the wobble in a way which depends both on the apparent luminosity and on the spatial properties of the source.
\par}
{
Efficient deconvolution of the Poisson statistics from the time{}-scale spectra (frequency domain) became possible after of the introduction the wavelet transform into studies of photon trains \citep{liszka99}. Using the wavelet transform as a pre{}-processing tool, non{}-linear modelling of the photon train \ variability may be performed using a neural network technique \citep{liszka00a,liszka03}. The problem of deconvolution of the spacecraft wobble may also be properly addressed using this technique.
\par}

\section{Wobble \textit{vs} Source Variability}
{
An advantage of the wobble is that it is known with a high degree of accuracy. That information may, for each observation period, be obtained from ROSAT archives. It may also be retrieved from photon history files, comparing photon sky coordinates and photon detector coordinates. As an example, both the x{}- (R.A.) and y{}-components (declination) of the wobble for ROR 701246 are shown in Fig.~\ref{fig1} as a function of time. The wobble is given in units of detector pixels. Since the wobble has a form of a symmetrical sawtooth signal, there will be a number of harmonics (see Table~\ref{table1}.) observed in the spectrum.
\par}
\clearpage
\begin{deluxetable}{lc}
\tablecaption{\label{table1}The wobble{}-related components.}
\tablewidth{0pt}
\tablehead{\colhead{Component} & \colhead{Period (sec)}}
\startdata
Fundamental & 400.0 \\
2:harmonics & 200.0 \\
3:harmonics & 133.3 \\
4:harmonics & 100.0 \\
5:harmonics & \phantom{0}80.0 \\
6:harmonics & \phantom{0}57.1 \\
\enddata
\end{deluxetable}
{
A detailed description of the instrument may be found in The ROSAT Users' Handbook. \ There are some other construction details that may generate some additional variability components. In the PSPC instrument, the window is supported by a system of two wire meshes: one coarse and one fine. It may be expected that both meshes will, during the wobble, generate a so{}-called moir\'e{}-pattern which enhances the 5:th harmonics of the wobble (80 sec). \ However, considering the diameters of the mesh wires, the obscuring scale size will be of the order of 10\textrm{"}. \ An obstacle of that size will not be very efficient in obscuring X{}-ray sources considering the limited resolving power of the PSPC instrument.
\par}
{
All harmonics, except for components caused by reflections in the instrument, must be phase{}-locked to the wobble. This property is important when it is necessary to remove the wobble{}-related components from the variability components intrinsic to the source. \ In the present work both the bulk of the Poisson component and the wobble{}-generated components are for the most part removed using a neural network model of spectral density in the time{}-scale \ {}- \ wavelet coefficient magnitude domain.
\par}
\begin{figure}[p]
\center
\includegraphics[width=.67\hsize,height=.4\hsize]{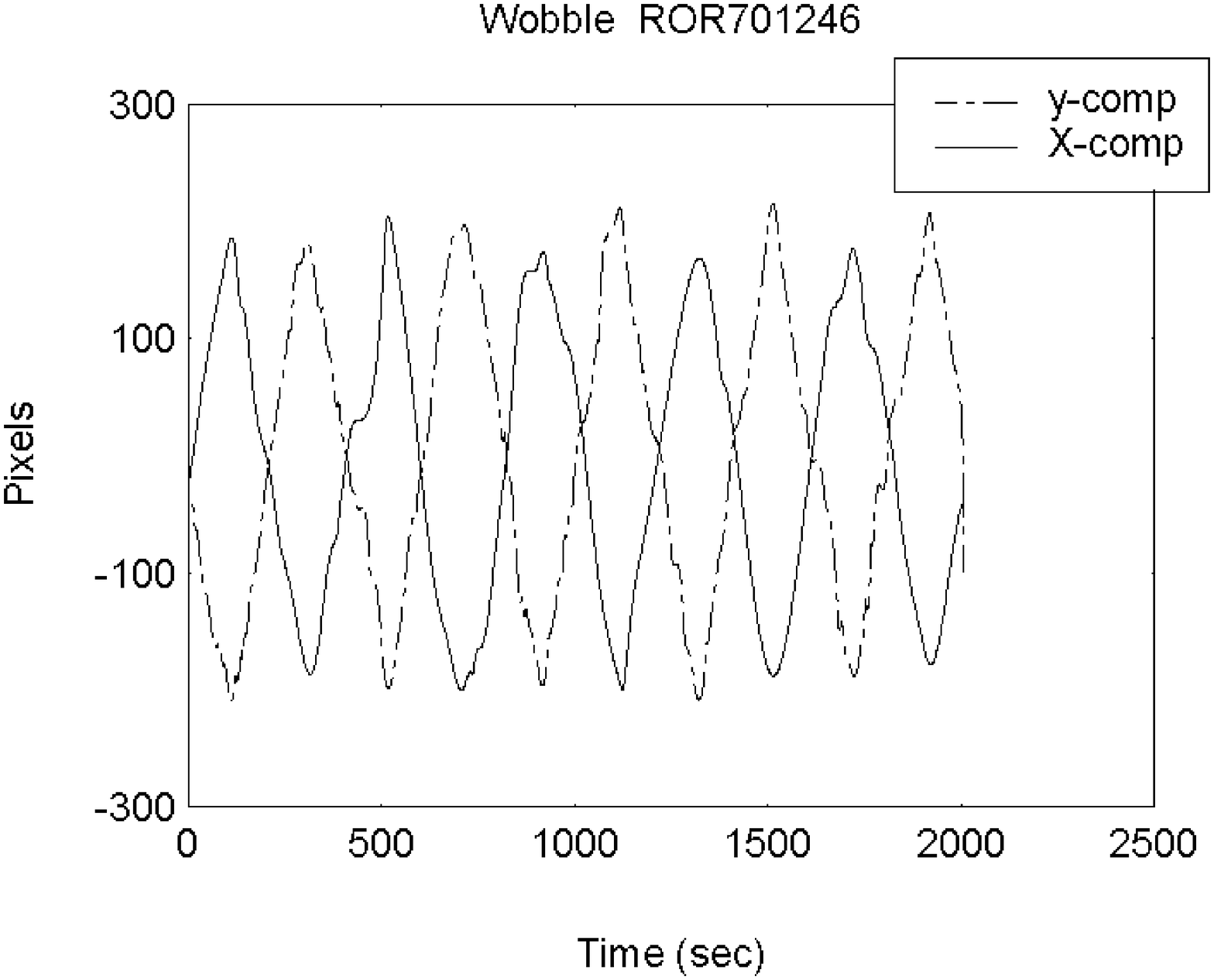}
\figcaption{X- and Y-component of the spacecraft wobble during ROR 701246 measured in units of detector
pixels.\label{fig2}}
\end{figure}
\section{FFT \textit{vs} Wavelet Spectrum}
{
In previous work \citep{liszka00b} a frequency component with a period of about 80 sec was shown as an example of the deterministic components in the variability of NGC 5548 during ROR 701246. The indicated period is close to one of the wobble{}-generated frequency components. Since the period was determined using the ``short time'' FFT (SFFT), it must be remembered the SFFT is not suitable for frequency analysis of the highly non{}-stationary photon counting rate.
\par}
\begin{figure}[tbp]
\center
\subfigure[]{\includegraphics[width=.47\hsize]{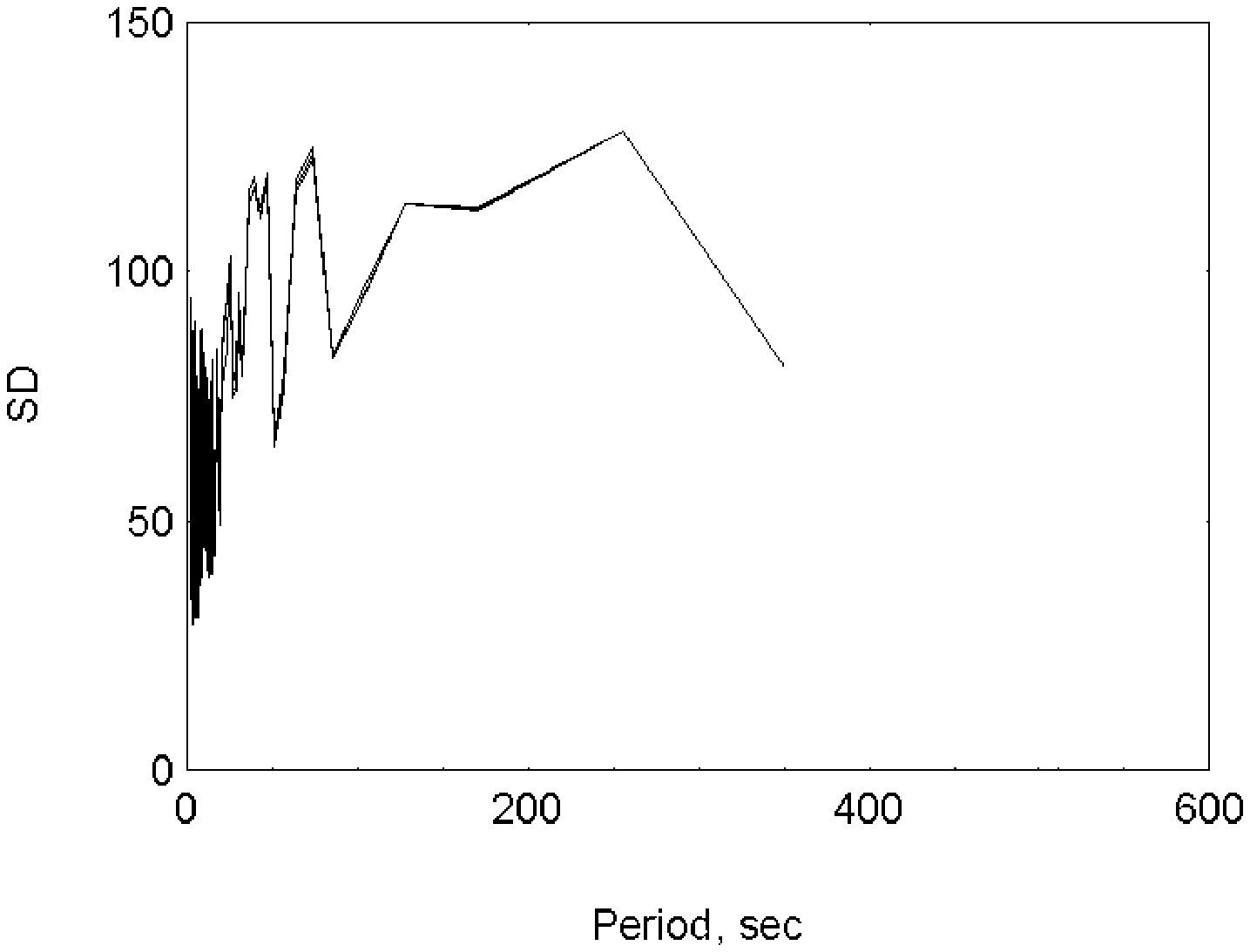}\label{fig3:a}}
\subfigure[]{\includegraphics[width=.47\hsize]{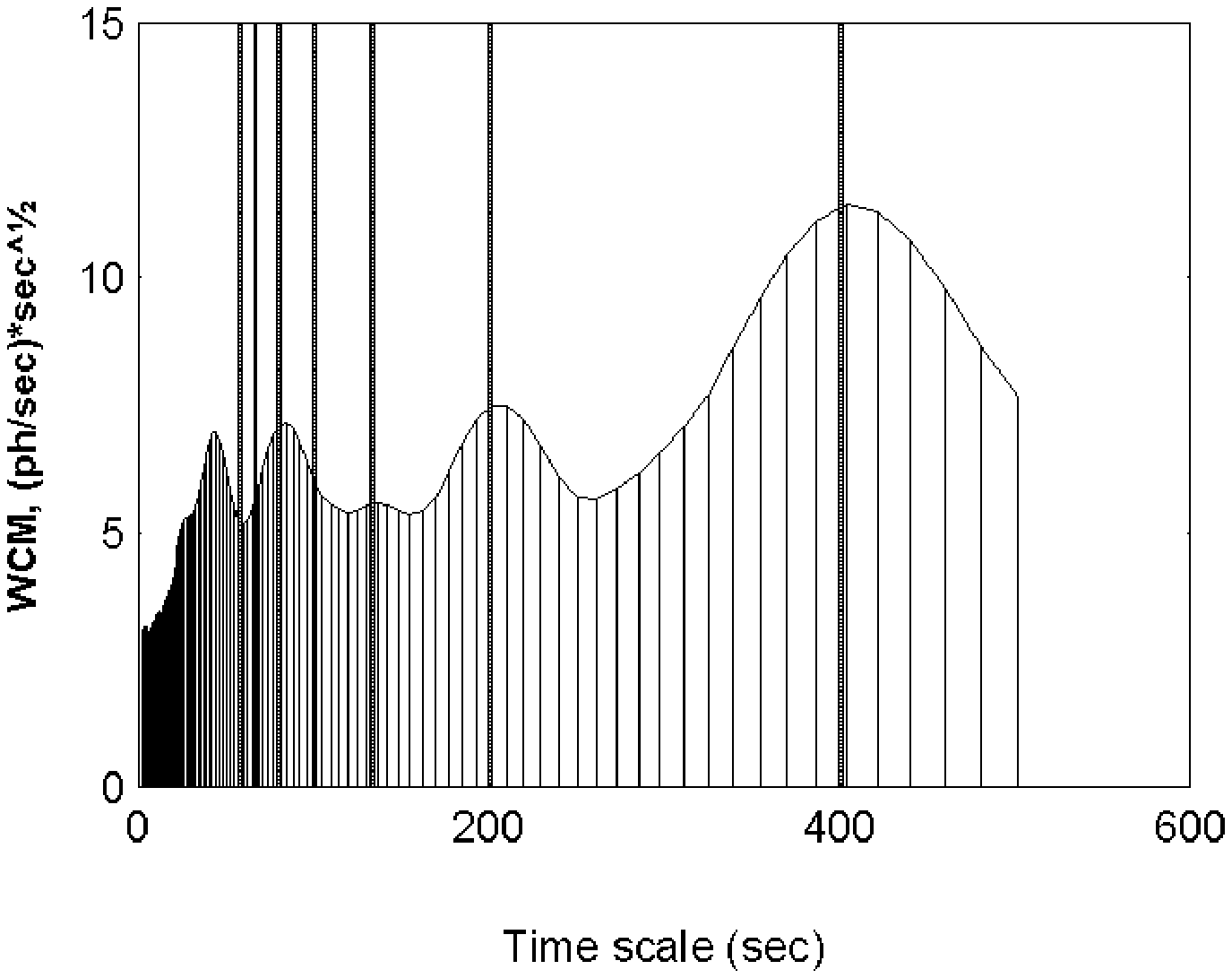}\label{fig3:b}}
\figcaption{The FFT spectral density (SD) of photon counting rate during ROR 701246, with 512 points in an analysing window plotted as a function of period (\subref{fig3:a}). The time averaged wavelet coefficient magnitude (WCM) of the original time series during 2007 sec of the observation period is shown in the right diagram. Positions of the wobble period and its harmonics are indicated by thick vertical lines.\label{fig3}}
\end{figure}
The FFT frequency spectrum is made using a constant width frequency interval. When plotted as a function of period (see Fig.~\ref{fig3}) the poor resolution in the low frequency part of the spectrum is obvious.

{
The wavelet spectrum (scalogram), shown in the right graph of Fig.~\ref{fig3}, uses a variable (frequency dependent) mother wavelet of Morlet type. It is also apparent that the resolution in the low frequency range of the spectrum is much better than in the FFT spectrum. It may be seen that the raw spectrum of photon counts seems to be dominated by the wobble frequency (400 sec period) and its harmonics.\par}
{In addition to more uniform frequency resolution there is an important advantage in using the wavelet spectrum. The scalogram may be converted into an ampligram \citep{liszka99} which is a set of inverse transforms of the results of non{}-linear filtering in the domain of wavelet coefficient magnitudes. For each band{}-pass of the wavelet coefficient magnitude the inverse transform will result in a time series corresponding to the signal within that magnitude interval. In the present study the wavelet coefficient magnitude (WCM) up to 40\% of its maximum value is analysed. A 4\% band{}-pass is shifted in 4\% steps obtaining 10 levels of the wavelet coefficient magnitude.\par}

\begin{figure}[ptb]
\center
\includegraphics[width=.47\hsize]{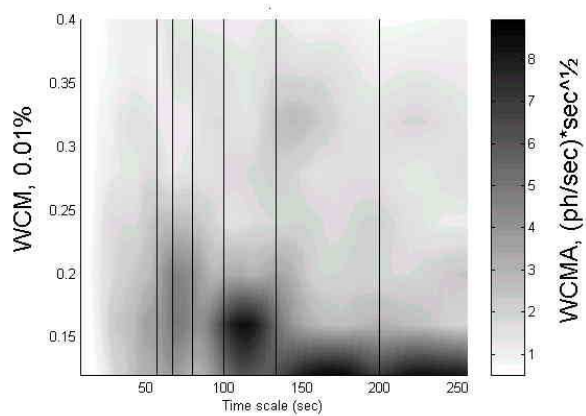}
\caption{The observed time scale spectrum for the entire observation interval of ROR 701246. Positions of the wobble and its harmonics are indicated by vertical lines. Gray scale gives the wavelet coefficient magnitude of the different levels in the ampligram, WCMA, \ (proportional to magnitude and persistence of structures). WCM is the wavelet coefficient magnitude of the original time series expressed in fractions of its maximum value.\label{fig4}}
\end{figure}

\begin{figure}[p]
%\subfigure[0--4\% WCM]{\includegraphics[width=.32\hsize,height=.27\hsize]{fig5a}\label{fig5:a}}%
%\subfigure[4--8\% WCM]{\includegraphics[width=.32\hsize,height=.27\hsize]{fig5b}\label{fig5:b}}%
%\subfigure[8--12\% WCM]{\includegraphics[width=.32\hsize,height=.27\hsize]{fig5c}\label{fig5:c}}
%\subfigure[12--16\% WCM]{\includegraphics[width=.32\hsize,height=.27\hsize]{fig5d}\label{fig5:d}}%
%\subfigure[16--20\% WCM]{\includegraphics[width=.32\hsize,height=.27\hsize]{fig5e}\label{fig5:e}}%
%\subfigure[20--24\% WCM]{\includegraphics[width=.32\hsize,height=.27\hsize]{fig5f}\label{fig5:f}}
%\subfigure[24--28\% WCM]{\includegraphics[width=.32\hsize,height=.27\hsize]{fig5g}\label{fig5:g}}%
%\subfigure[28--32\% WCM]{\includegraphics[width=.32\hsize,height=.27\hsize]{fig5h}\label{fig5:h}}%
%\subfigure[32--36\% WCM]{\includegraphics[width=.32\hsize,height=.27\hsize]{fig5i}\label{fig5:i}}
%\subfigure[36--40\% WCM]{\includegraphics[width=.32\hsize,height=.27\hsize]{fig5j}\label{fig5:j}}
\includegraphics[width=\hsize]{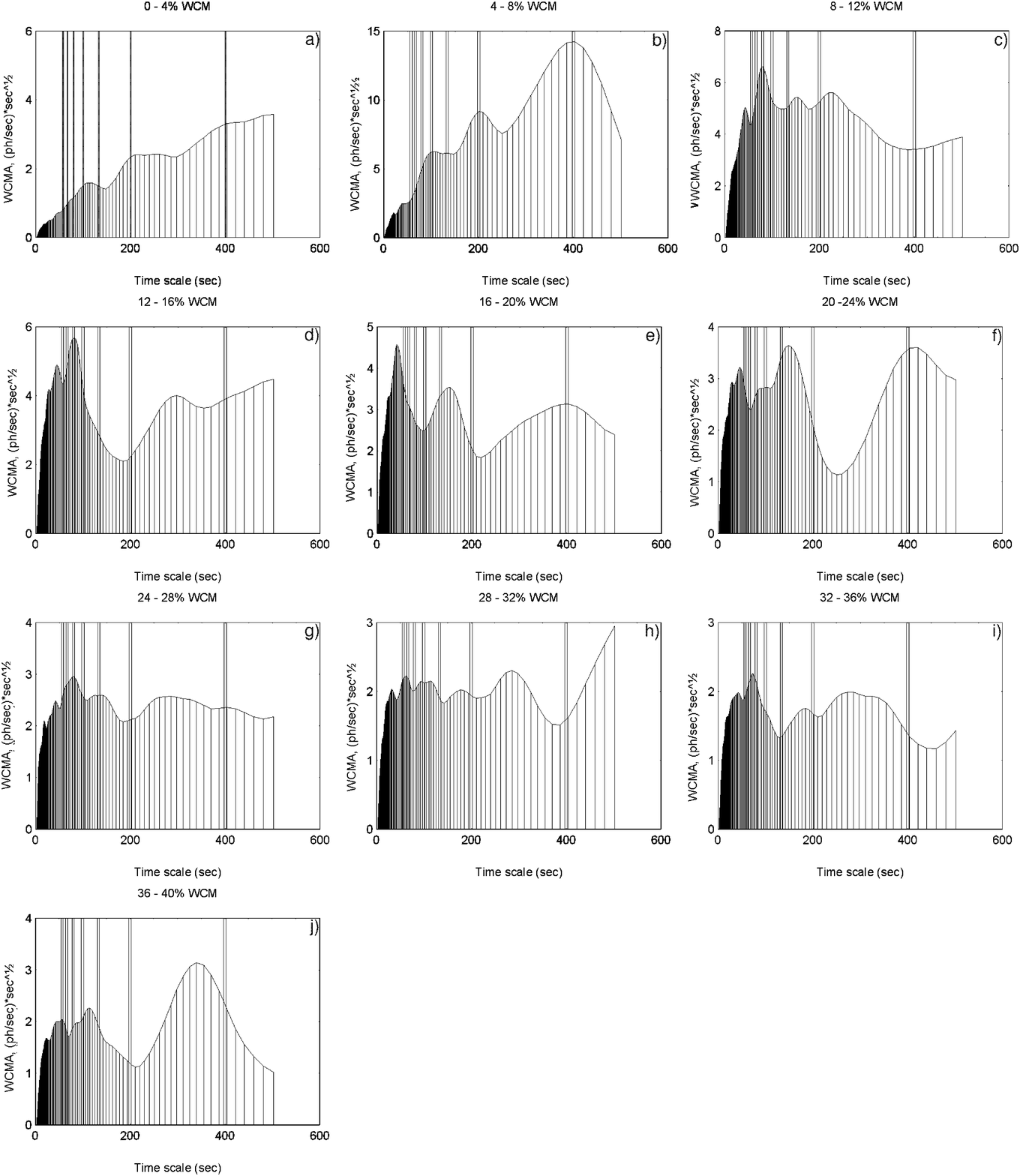}
\figcaption{The time scale spectrum for the entire observation interval of ROR 701246 shown in the form of individual graphs a {--} j, one for each 4\% level of the wavelet coefficient magnitude of the original time series. \ Positions of the wobble and its harmonics are indicated by double vertical bars.\label{fig5}}
\end{figure}

{The present technique gives the possibility of decomposing a signal consisting of components of different magnitudes. The method was developed to separate the Poisson variability from intrinsic variations in counting rate. \ In the same way, it will be shown that the combined effect of the Poisson distribution and of the spacecraft wobble may be removed from the observed counting rate. Another forward wavelet transform applied to the ampligram converts it into the time scale spectrum \citep{liszka99}, showing the wavelet coefficient magnitude for each ampligram level (WCMA) as a function of time scale. The time scale spectrum (TSS) is thus the spectrum of time scale at different levels of the wavelet coefficient magnitude of the original time series, averaged over the entire observation period. The WCMA is proportional to the persistence (cf. section 6) of a component with a given magnitude and is represented by the darkness of the gray scale in Figs.~\ref{fig4}, \ref{fig5} and \ref{fig6}. The time scale spectrum for the entire observation interval of ROR 701246 is shown in Fig.~\ref{fig4} as a 3{}-D plot. Positions of the wobble and its harmonics are indicated by vertical lines.\par}

The time scale spectrum of Fig.~\ref{fig4} may be displayed in the form of 10 \ 2{}-D graphs. Position of the wobble and its harmonics is indicated by double vertical bars on each sub{}-diagram.

%\begin{figure}[tbp]
%\subfigure[0--4\% WCM]{\includegraphics[width=.47\hsize,height=.27\hsize]{fig5a}\label{fig5:a}}%
%\subfigure[4--8\% WCM]{\includegraphics[width=.47\hsize,height=.27\hsize]{fig5b}\label{fig5:b}}
%\subfigure[8--12\% WCM]{\includegraphics[width=.47\hsize,height=.27\hsize]{fig5c}\label{fig5:c}}%
%\subfigure[12--16\% WCM]{\includegraphics[width=.47\hsize,height=.27\hsize]{fig5d}\label{fig5:d}}
%\subfigure[16--20\% WCM]{\includegraphics[width=.47\hsize,height=.27\hsize]{fig5e}\label{fig5:e}}%
%\subfigure[20--24\% WCM]{\includegraphics[width=.47\hsize,height=.27\hsize]{fig5f}\label{fig5:f}}
%\caption{The time scale spectrum for the entire observation interval of ROR 701246 shown in the form of individual graphs \subref{fig5:a} {--} \subref{fig5:j}, one for each 4\% level of the wavelet coefficient magnitude of the original time series. \ Positions of the wobble and its harmonics are indicated by double vertical bars.\label{fig5}}
%\end{figure}
%\begin{figure}[tbp]
%\addtocounter{figure}{-1}
%\addtocounter{subfigure}{6}
%\subfigure[24--28\% WCM]{\includegraphics[width=.47\hsize,height=.27\hsize]{fig5g}\label{fig5:g}}%
%\subfigure[28--32\% WCM]{\includegraphics[width=.47\hsize,height=.27\hsize]{fig5h}\label{fig5:h}}
%\subfigure[32--36\% WCM]{\includegraphics[width=.47\hsize,height=.27\hsize]{fig5i}\label{fig5:i}}%
%\subfigure[36--40\% WCM]{\includegraphics[width=.47\hsize,height=.27\hsize]{fig5j}\label{fig5:j}}
%\addtocounter{figure}{1}% need to be before \contcaption
%\setcounter{subfigure}{0}
%\contcaption{Continued}
%\end{figure}
{
It may be seen that the strongest coincidence between the wobble and its second harmonics (200 sec) is in the 8\% band (Fig~\ref{fig5}b). The 80 sec peak, expected in the wobble{}-related variability, coincides with an observed peak only in the 12\% and 16\%, 4\%{}-pass{}-bands (Figs~\ref{fig5}c,d). \ At all higher levels of the wavelet coefficient magnitude peaks occurring in the spectrum cannot be correlated with either the wobble, or with its harmonics.
\par}

\section{Quantifying the Influence of the Wobble on the Counting Rate Variability}
{The fact that a spectral peak coincides with a given wobble{}-related frequency cannot be considered as a proof of its origin. If a signal component and the corresponding spectral peak is really wobble{}-related, there must be a well{}-defined phase relation between the wobble and the signal. However, this will not be the case for components generated by reflections in the instrument. A simple method to test the phase relation is to calculate the cross{}-correlation function between the wobble and all 10 levels of the counting rate ampligram. The cross{}-correlation function (CCF) was calculated using the commercial software package SPSS. Results of calculations are shown in Table~\ref{table2} where the peak cross{}-correlation is listed together with the corresponding phase difference (lag time) and confidence limits.\par}
\clearpage
\begin{deluxetable}{ccccc}
\tablecaption{\label{table2}Cross{}-correlation (CCF) between the spacecraft wobble and different ampligram ranges (wavelet coefficient magnitudes) for ROR 701246}
\tablewidth{0pt}
\tablehead{
\colhead{Graph no.} &
\colhead{WCM range (\%)} &
\colhead{Phase diff.(deg)} &
\colhead{Peak CCF} &
\colhead{Confidence limits +/{}-}
}
\startdata
\ \ref{fig5}a &  \phantom{0}0 {}- \phantom{0}4 &   $\phantom{-}29$ &  $\phantom{-}0.250$ & 0.023 \\
\ \ref{fig5}b &  \phantom{0}4 {}- \phantom{0}8 &   $\phantom{-}32$ &  $\phantom{-}0.343$ & 0.023 \\
\ \ref{fig5}c &  \phantom{0}8 {}-           12 &             $-81$ &            $-0.103$ & 0.023 \\
\ \ref{fig5}d &            12 {}-           16 &             $-39$ &  $\phantom{-}0.083$ & 0.023 \\
\ \ref{fig5}e &            16 {}-           20 &             $-32$ &            $-0.067$ & 0.023 \\
\ \ref{fig5}f &            20 {}-           24 &             $-13$ &  $\phantom{-}0.079$ & 0.022 \\
\ \ref{fig5}g &            24 {}-           28 &   $\phantom{ }-2$ &            $-0.052$ & 0.022 \\
\ \ref{fig5}h &            28 {}-           32 &             $-59$ &  $\phantom{-}0.026$ & 0.023 \\
\ \ref{fig5}i &            32 {}-           36 &   $\phantom{-}37$ &  $\phantom{-}0.036$ & 0.023 \\
\ \ref{fig5}j &            36 {}-           40 &             $-95$ &            $-0.059$ & 0.023 \\
\enddata
\end{deluxetable}
{
It may be seen that the largest value of CCF is obtained for the wavelet coefficient magnitude interval between 4 and 8\%, the same interval where the largest 400 and 200 sec peaks are observed (Fig.~\ref{fig5}b). However, the observed average phase difference (32 degrees) cannot be explained by assuming that both dominating components are wobble{}-related. All calculated peak values of CCF are outside the confidence limits.
\par}
\subsection{Modelling of Variability in the Time Scale Domain}
{
Now we will analyse the effect of wobble and stochastic Poisson statistics on the resulting overall X{}-ray variability of the observed sources. Since the analyzed counting rates are small and have an extremely skew distribution with respect to apparent luminosity (dominance of low average photon counting rates), a conventional averaging process would give very little information. Photon history files for NGC 5548 were converted into a photon counting rate using 1{}-sec sampling bins. \ Photon counting rate files were converted into ampligrams \citep{liszka99}, and subsequently into time scale spectra. The time scale spectra covered the interval of wavelet coefficient magnitudes up to 40\% of the maximum magnitude and the time scales between 5 and 256 seconds. \ The upper time scale limit was now shorter than in the case of ROR 701246 displayed in Figs.~\ref{fig4} and \ref{fig5}, since the model (described below) was based on all observation periods of NGC 5548, some of them with only 1000 sec duration. Each time scale spectrum was a 10 x 128 matrix (number of levels in the ampligram x number of dilations in the wavelet transform).
\par}
{
Since the amount of data was not large enough to properly remove the dependence on apparent luminosity using conventional statistical methods, an alternative method has been used. \ A back{}-propagation (BP) neural network was used for modelling the available data. \ Training was performed using 85\% of the entire data while the remaining 15\% was used to test the model. \ The training data was randomly selected. \ The model consisted of the following variables:
\par}

\paragraph{Input}
\begin{itemize}
\item{Average counts (apparent luminosity), which were determined for each observation period.\par}
\item{Photon energy index indicating the energy interval\par}
\item{M {--} wavelet coefficient magnitude in \% of its maximum value. 20 levels between 0 and 40\% are used in the present study.\par}
\end{itemize}
\paragraph{Output}
{
A 128{}-component spectral vector being that row of time scale spectrum matrix which corresponds to the actual value of M.
\par}
{
The BP neural network used to model the data consists of 3 processing elements (PE) in the input layer, 128 PE in the output layer and 130 PE in the hidden layer. A recall of the trained network using the test data (not used for training) gives one value of the correlation coefficient between the pair of spectral vectors: that retrieved from the model and the measured one. One value of the correlation coefficient is obtained for each component of the spectral vector. That set of correlation coefficients is a measure of both the predictive ability of the model and also its generalizing ability (response to previously unseen data). \ Values of correlation coefficients for the present network lie between 0.8 and 0.95 in the entire range of time scales. These very high correlation coefficients indicate the very high quality of the present model and most likely the high consistency of the data used to generate the model.
\par}
{
Results using the NGC 5548 model for the data of ROR 701246 for the photon energies E{\textless}0.5 keV are presented below. The top diagram of Fig.~\ref{fig6} shows: the observed time scale spectrum for the data of ROR 701246. The middle diagram shows the time scale spectrum retrieved from the model. That spectrum contains apparent luminosity dependent components present in all NGC5548 data, i.~e, the wobble{}-related and Poisson{}-related components. The difference between the above spectra may be attributed to intrinsic variations involving ``elementary events'' in the source during that particular period of time. The difference spectrum, exceeding error intervals, is shown in the bottom graph of Fig.~\ref{fig6}.
\par}
{
Of course, this method will only pick out transient deterministic structures {--} any structure, such as a periodicity, which is common to all observation periods will also be eliminated by this analysis. At the same time, one may speculate that perhaps certain wobble{}-related harmonic features do not show up in all observation periods, and thus are not excluded by this averaged subtraction process. \ They may only appear when the source is nearly on{}-axis, for instance. This is indeed possible {--} it is also clear, for instance, that the \~{}80 s harmonics of the wobble are only apparent in three of the pass{}-bands of Fig.~\ref{fig5}a, b and h. However, the fact the detected structures are not exactly at the harmonic frequency and are not phase{}-locked to the wobble indicates that its origin is elsewhere. Furthermore, the presence of persistent quasi{}-periodicities at frequencies different from those of the wobble harmonics is evidence that a process intrinsic to the source is at work. If the quasi{}-periodic structures were only at or near harmonic frequencies, that would be suspicious.
\par}
\begin{figure}[p]
\center
\subfigure[]{\includegraphics[width=.47\hsize]{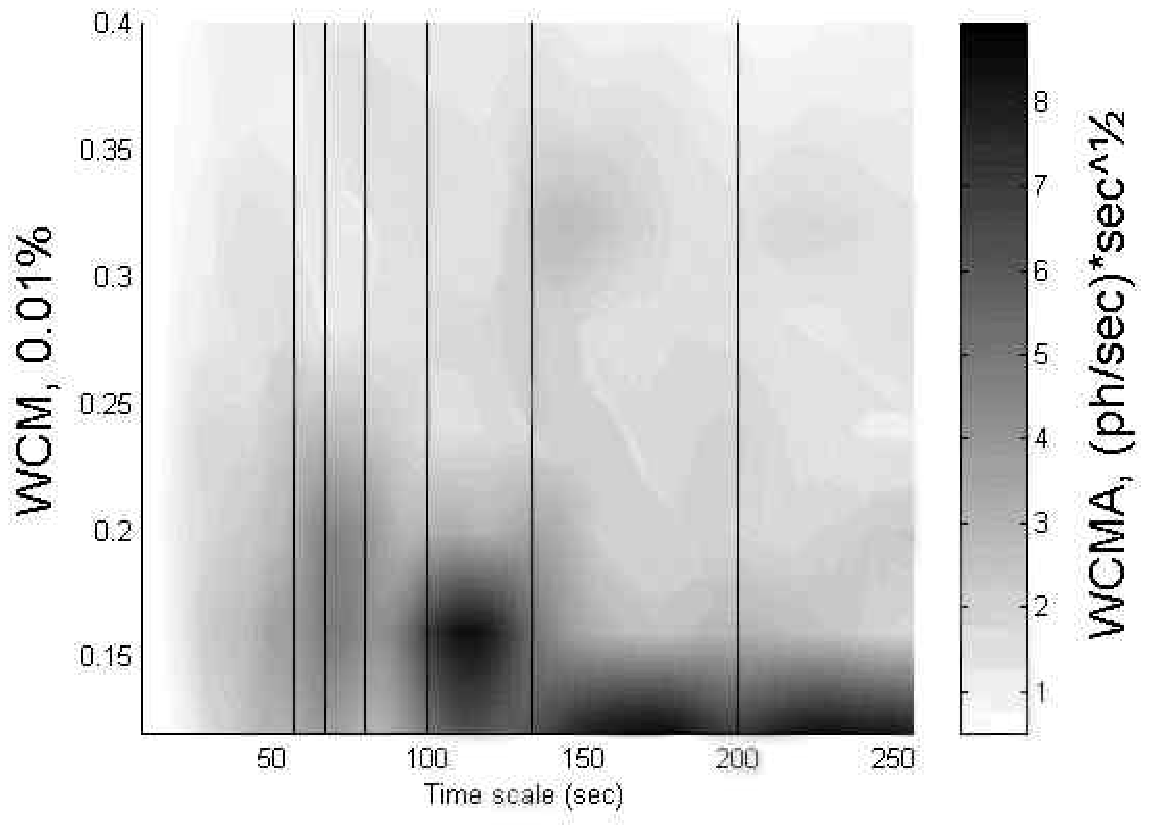}\label{fig6:a}}%
\subfigure[]{\includegraphics[width=.47\hsize]{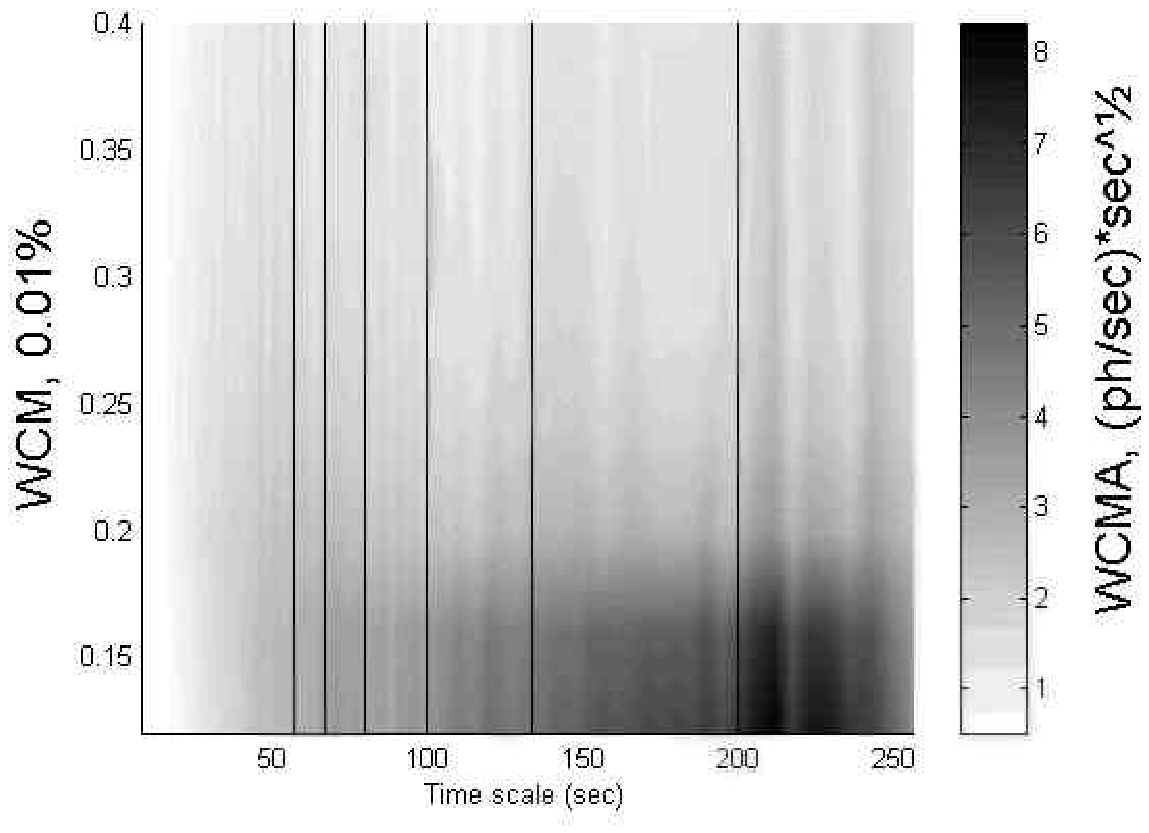}\label{fig6:b}}
\subfigure[]{\includegraphics[width=.47\hsize]{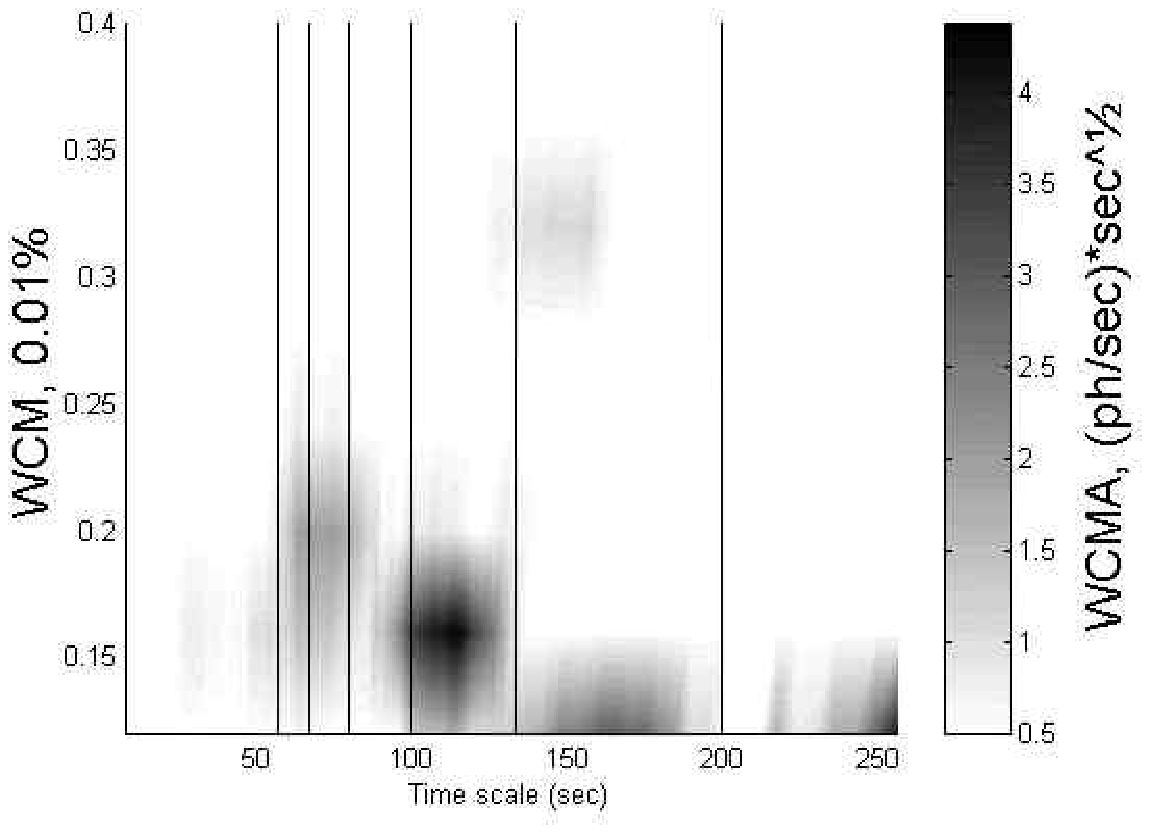}\label{fig6:c}}
\caption{Time scale spectra of NGC 5548, ROR 701246, \subref{fig6:a} observed time scale spectrum. \subref{fig6:b} ``neural network averaged'' time scale spectrum retrieved from the model. \subref{fig6:c} difference between top and middle spectrum (called here the difference spectrum). Only differences exceeding the average error interval of 0.05 are plotted. \ Wobble harmonics are indicated by vertical lines.\label{fig6}}
\end{figure}
{
In NGC 5548 the spectrum retrieved from the model for the actual counting rate is significantly different from the observed spectrum. \ As it was mentioned at the beginning of this section, the model extracts from the observed data common features present in the data. The obvious common features are the Poisson statistics, the apparent luminosity dependence and wobble harmonics. The residuals after subtraction of the model from the observed NGC5548 spectra seem to consist of a number of deterministic events, limited in time scale (frequency) and magnitude. As we have pointed out in earlier papers, deterministic components of the variability appear as vertical or circular features in these time scale spectra, whereas stochastic components appear as prominent horizontal features. As we see, the components resulting from this subtraction do have circular shape, or vertical extent.
\par}
{
These structures, or elementary events are mainly limited to a single interval of wavelet coefficient magnitude (4\% bandwidth) and to a narrow band of time scale indicating a stable, nearly monochromatic signal with a duration (persistence) proportional to its intensity. \ Figure~\ref{fig7} shows a sample of difference spectra from different observation periods of NGC 5548.
\par}
\begin{figure}[p]
\subfigure[ROR 701238]{\includegraphics[width=.47\hsize,height=.28\hsize]{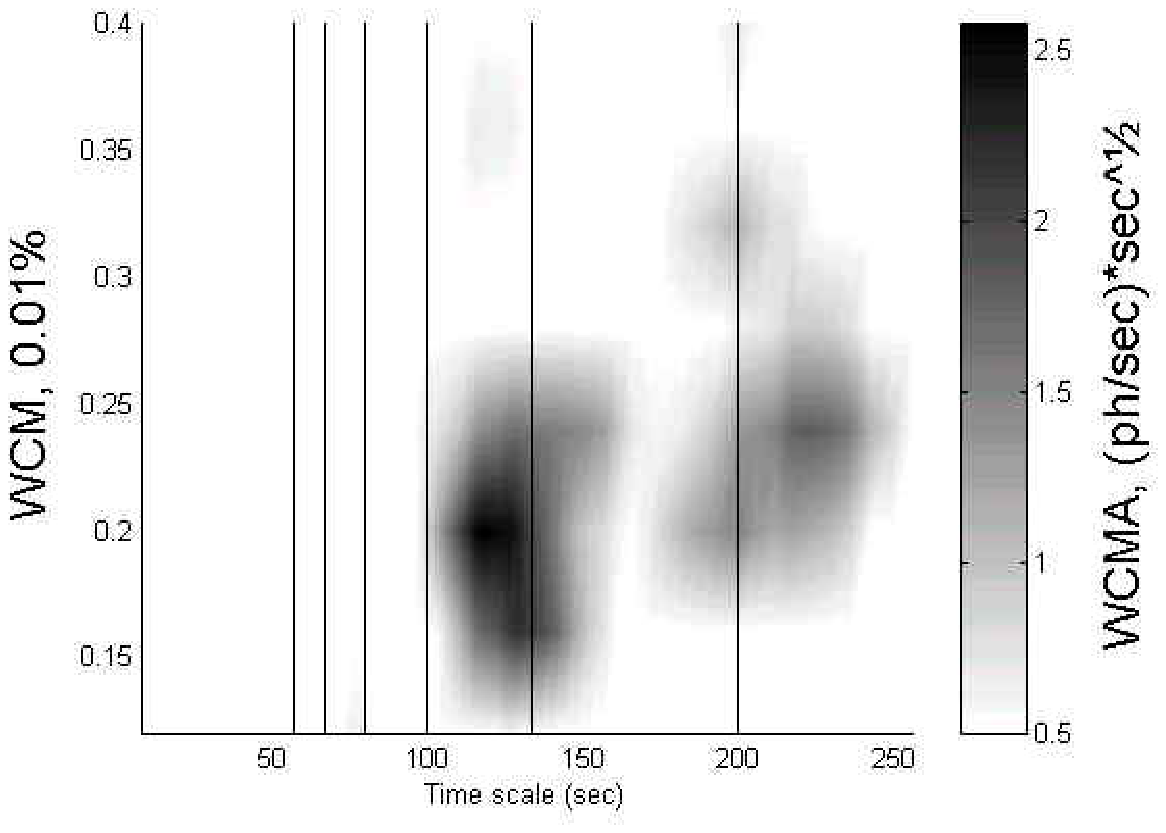}\label{fig7:a}}%
\subfigure[ROR 701239]{\includegraphics[width=.47\hsize,height=.28\hsize]{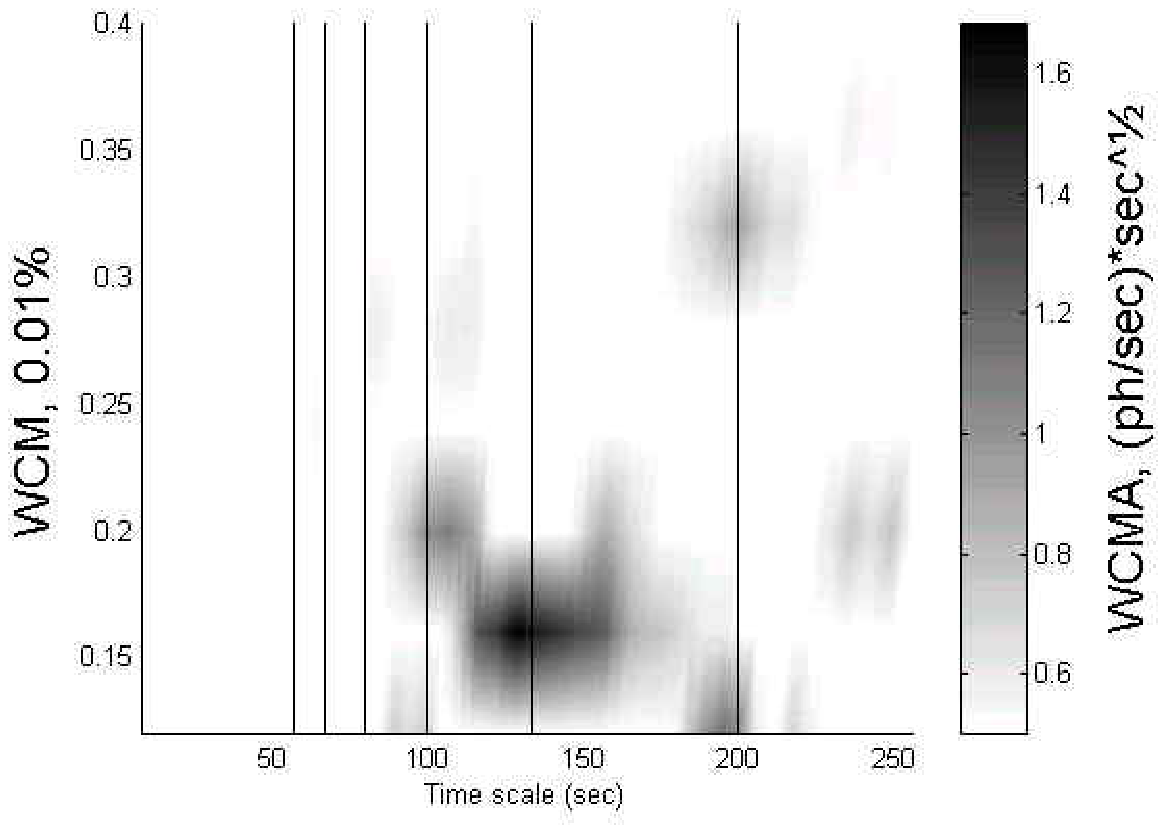}\label{fig7:b}}
\subfigure[ROR 701241]{\includegraphics[width=.47\hsize,height=.28\hsize]{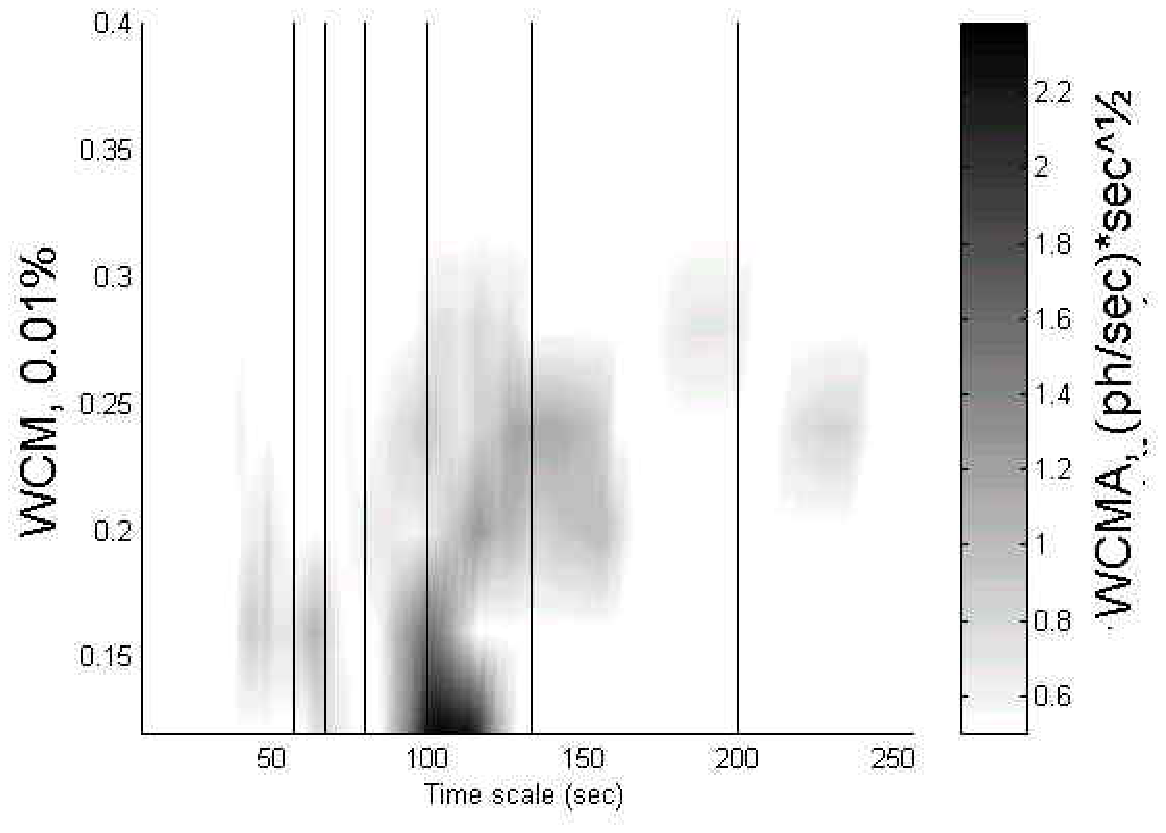}\label{fig7:c}}%
\subfigure[ROR 701248]{\includegraphics[width=.47\hsize,height=.28\hsize]{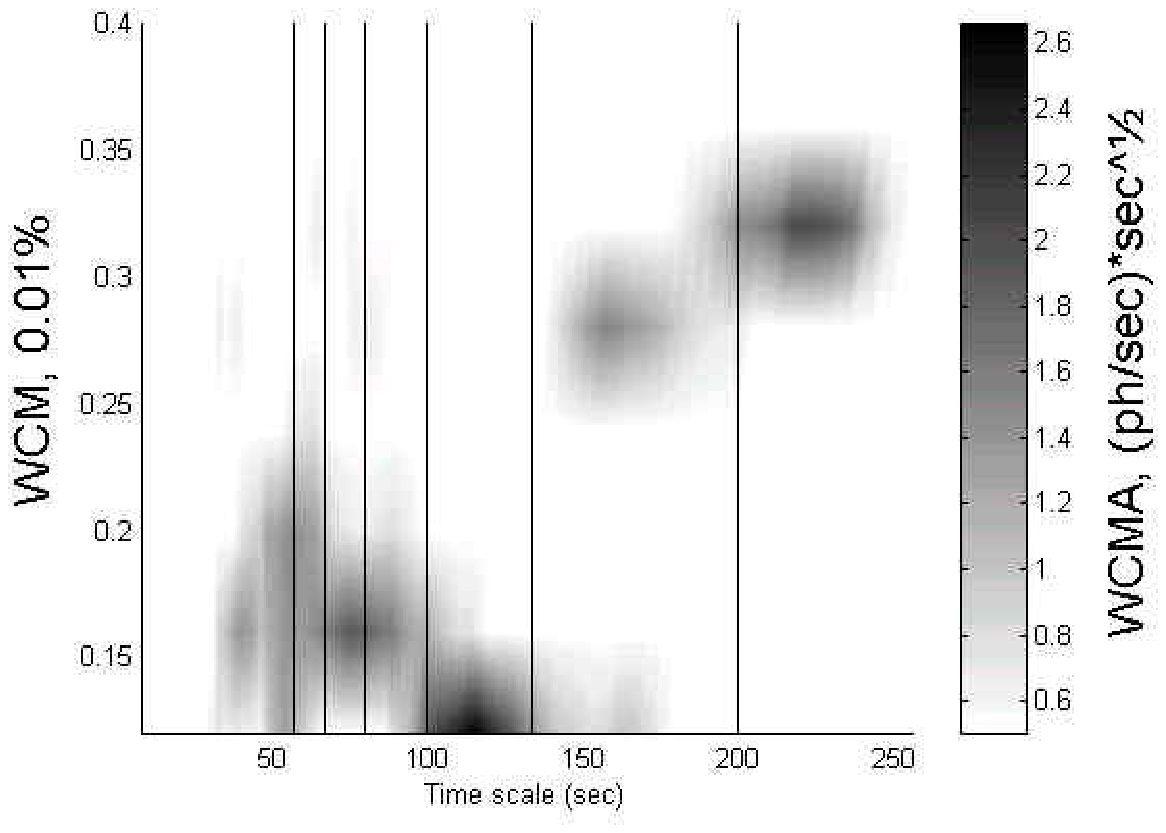}\label{fig7:d}}
\subfigure[ROR 701270]{\includegraphics[width=.47\hsize,height=.28\hsize]{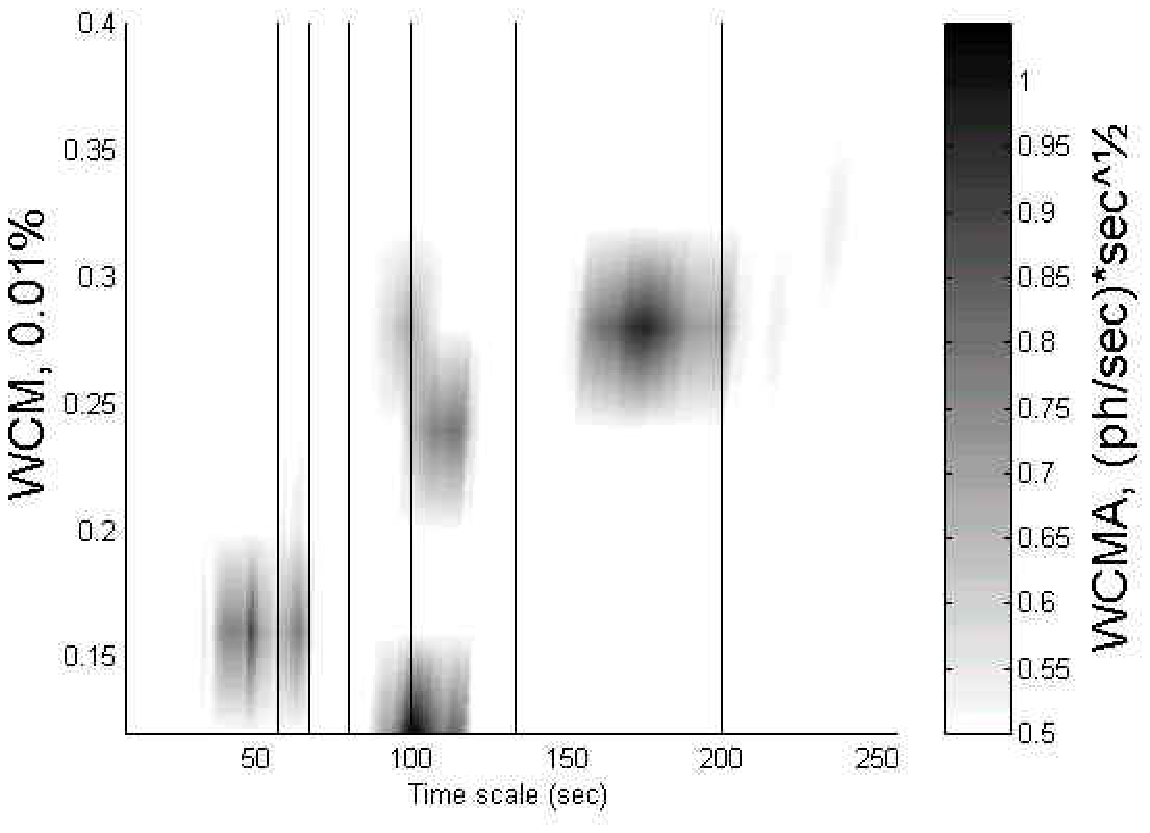}\label{fig7:e}}%
\subfigure[ROR 701271]{\includegraphics[width=.47\hsize,height=.28\hsize]{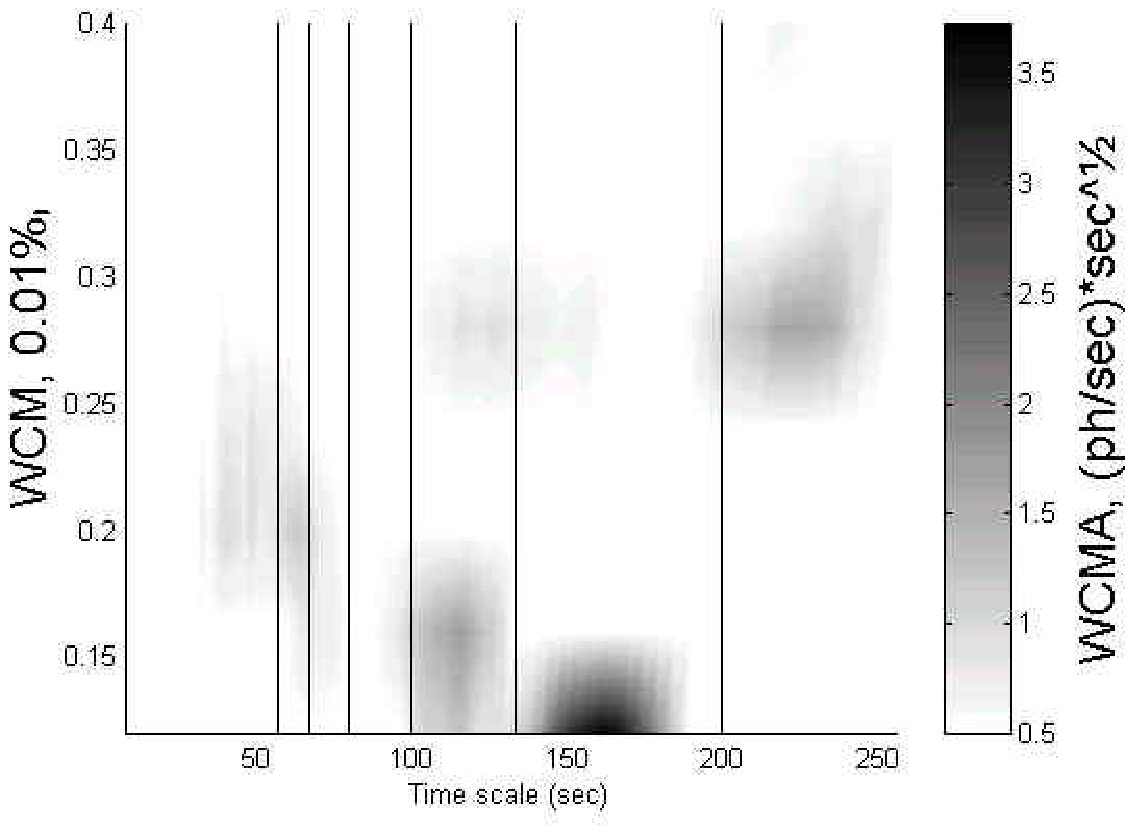}\label{fig7:f}}
\caption{A sample of typical difference spectra (observed time scale spectrum {--} the spectrum retrieved from the model) for NGC 5548. Wobble harmonics are indicated by vertical lines.\label{fig7}}
\end{figure}
{
The difference spectra were calculated for 30 observation periods of NGC 5548. All those periods were longer than 1024 seconds. \ All events were identified automatically and their positions in the spectrum were read out. The entire time scale spectrum (wavelet coefficient magnitudes 0 {--} 40\%, time scales 5 {--} 256 seconds) was divided into 10 x 16 equal bins and all events were sorted into corresponding bins. The result of event sorting is shown in Fig.~\ref{fig8:a} (average intensity) and ~\subref{fig8:b} (number of events within a bin). \ The result of the statistics shows that the events may occur anywhere within the investigated range of time scale spectrum. There is a slight increase in intensity towards small wavelet coefficient magnitudes, which could indicate that weak events are more persistent.
\par}
\begin{figure}[p]
\subfigure[]{\includegraphics[width=.47\hsize]{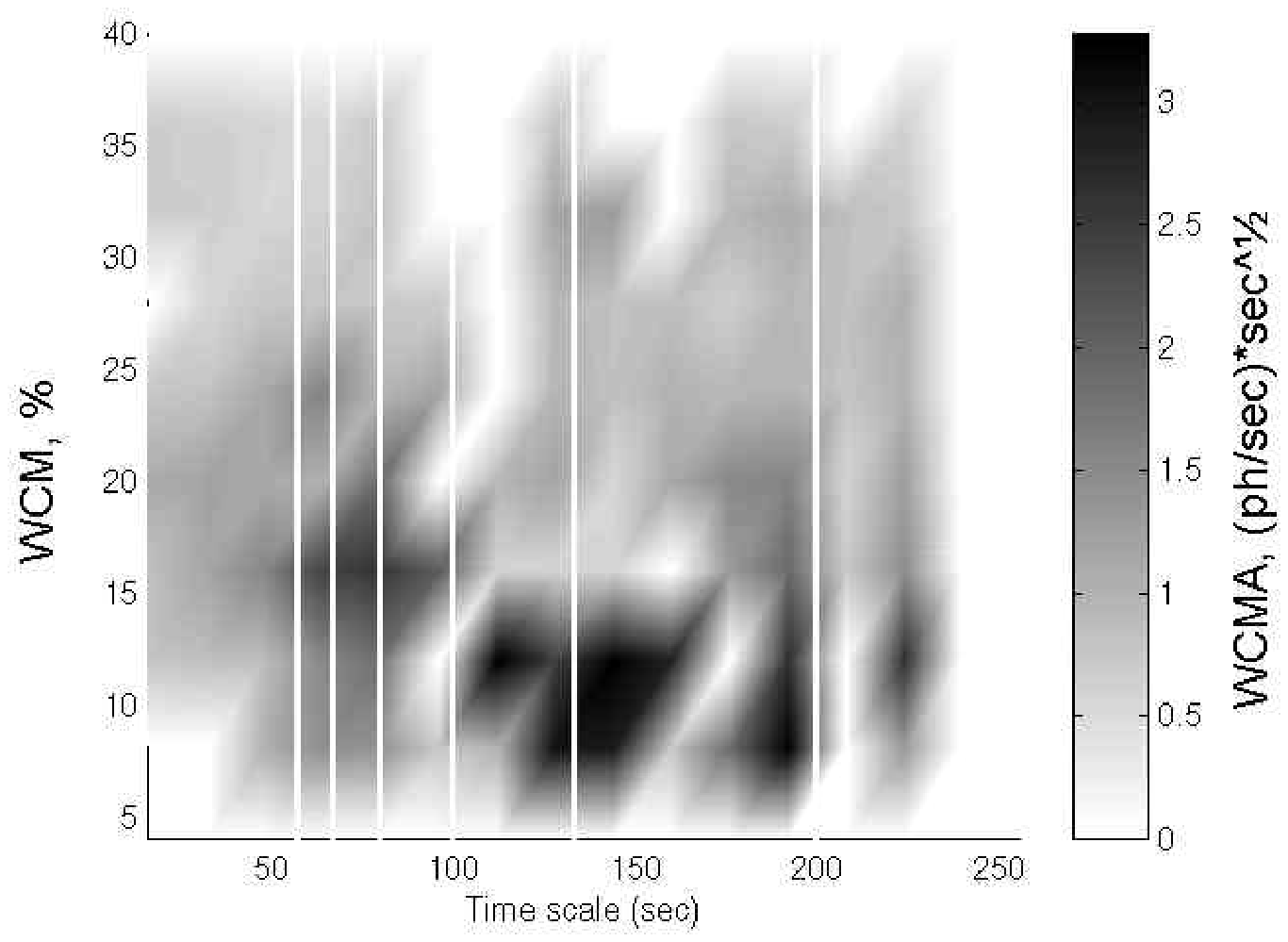}\label{fig8:a}}%
\subfigure[]{\includegraphics[width=.47\hsize]{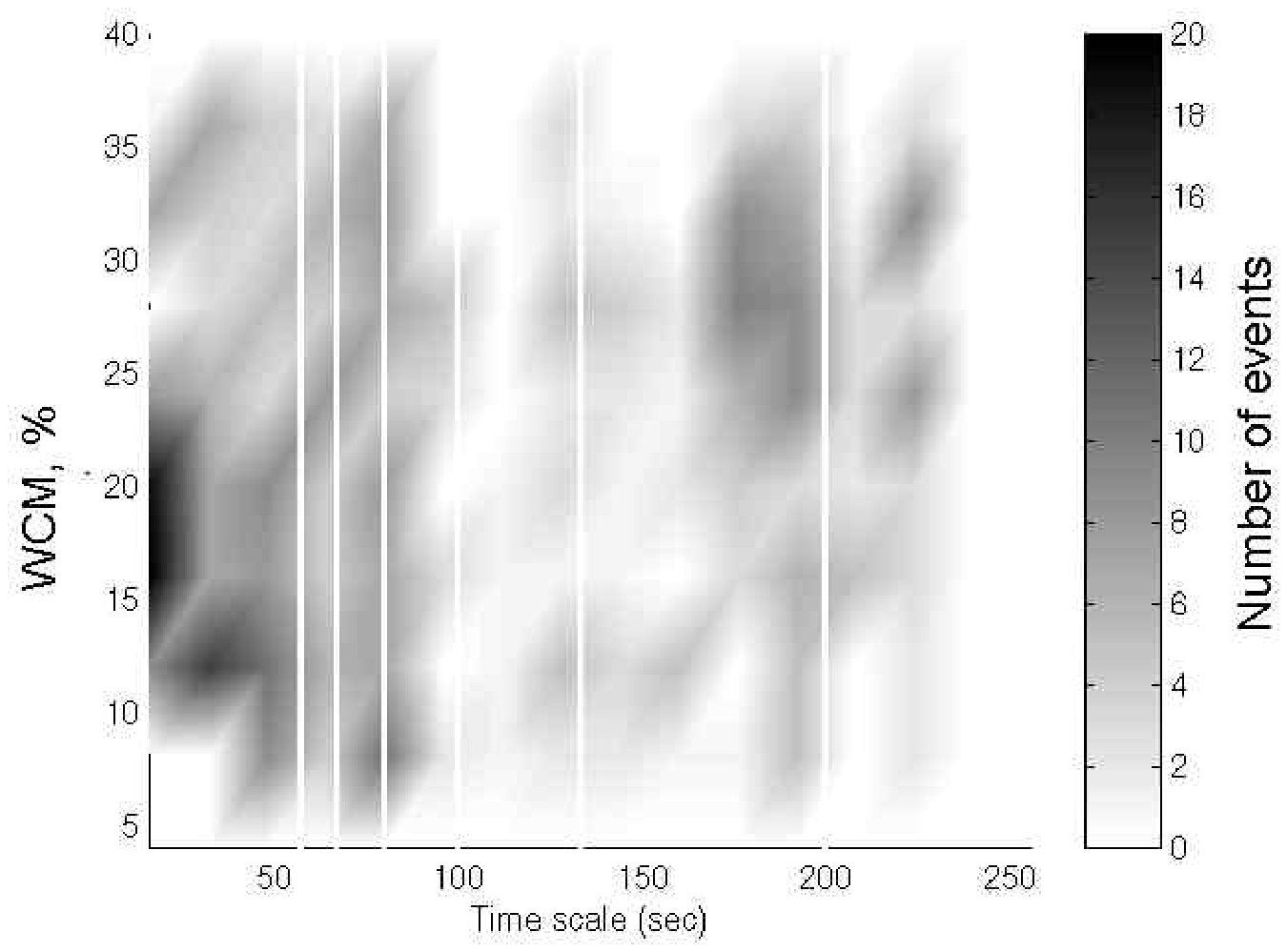}\label{fig8:b}}
\caption{Average intensity of the elementary events (left graph) and the number of events per bin (right graph) during 30 observation periods of NGC 5548. Wobble harmonics are indicated by vertical lines.}
\label{fig8}
\end{figure}
\bigskip
{
There is an interesting feature that may be seen in the graphs of Fig.~\ref{fig8}. Both the average intensity (persistence) of events and their number seem to be suppressed at both 100 and 200 sec scales, which correspond to the second and third harmonics of the wobble. The cut{}-off at 230 sec is a result of the method used to detect the events. That means that the wobble suppresses the visibility of elementary events in the time scale spectrum, rather than enhances them. In subtracting out Poisson statistics and wobble components much weaker elementary events at the wobble harmonics frequencies are at least partially removed.
\par}
\section{Persistence}
\begin{figure}[p]
\centering
\includegraphics[width=.57\hsize,height=.27\hsize]{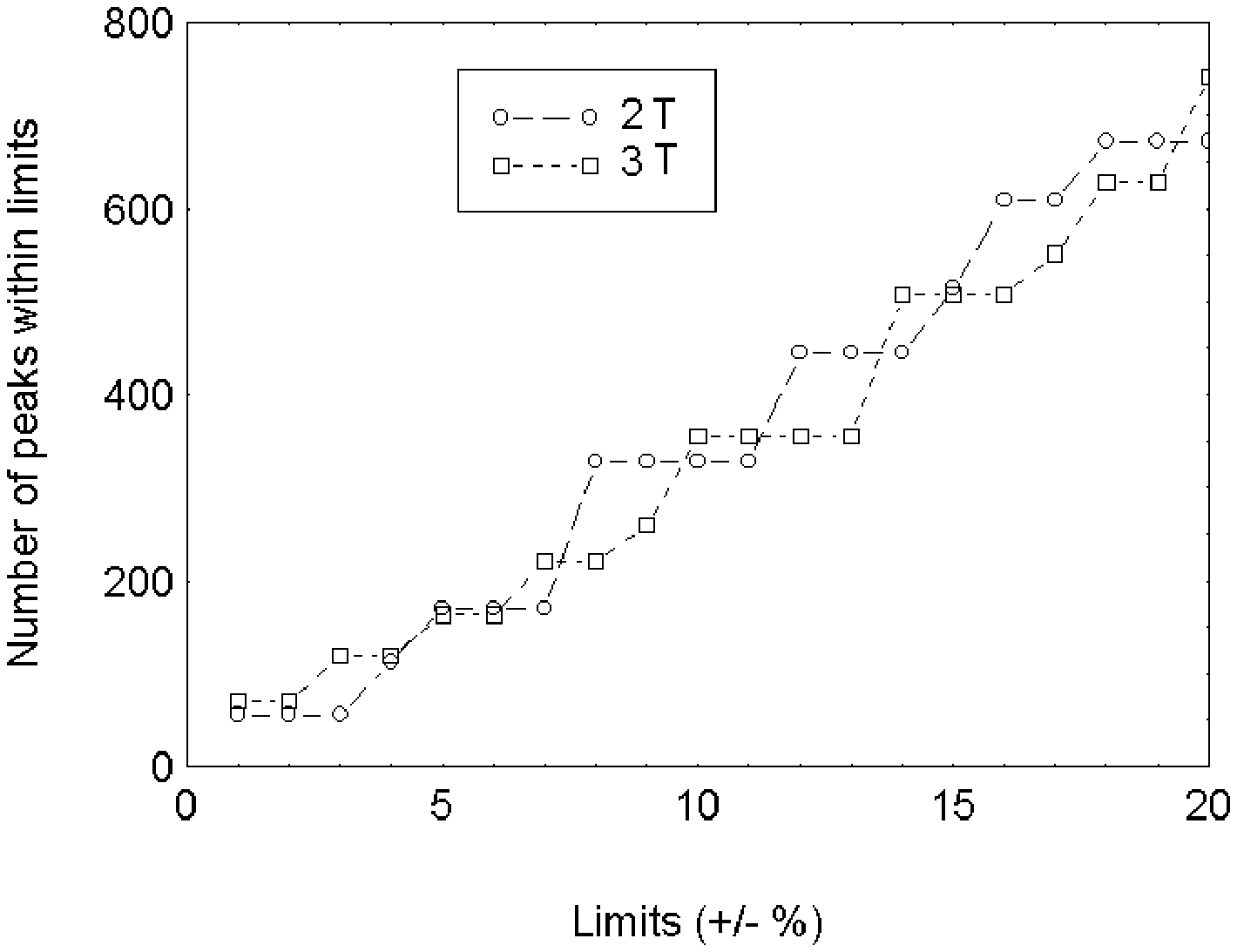}
\caption{Number of peaks identified as multiples when the acceptance limits increase\label{fig9}}
\end{figure}

\begin{figure}[p]
\centering
\includegraphics[width=.57\hsize,height=.27\hsize]{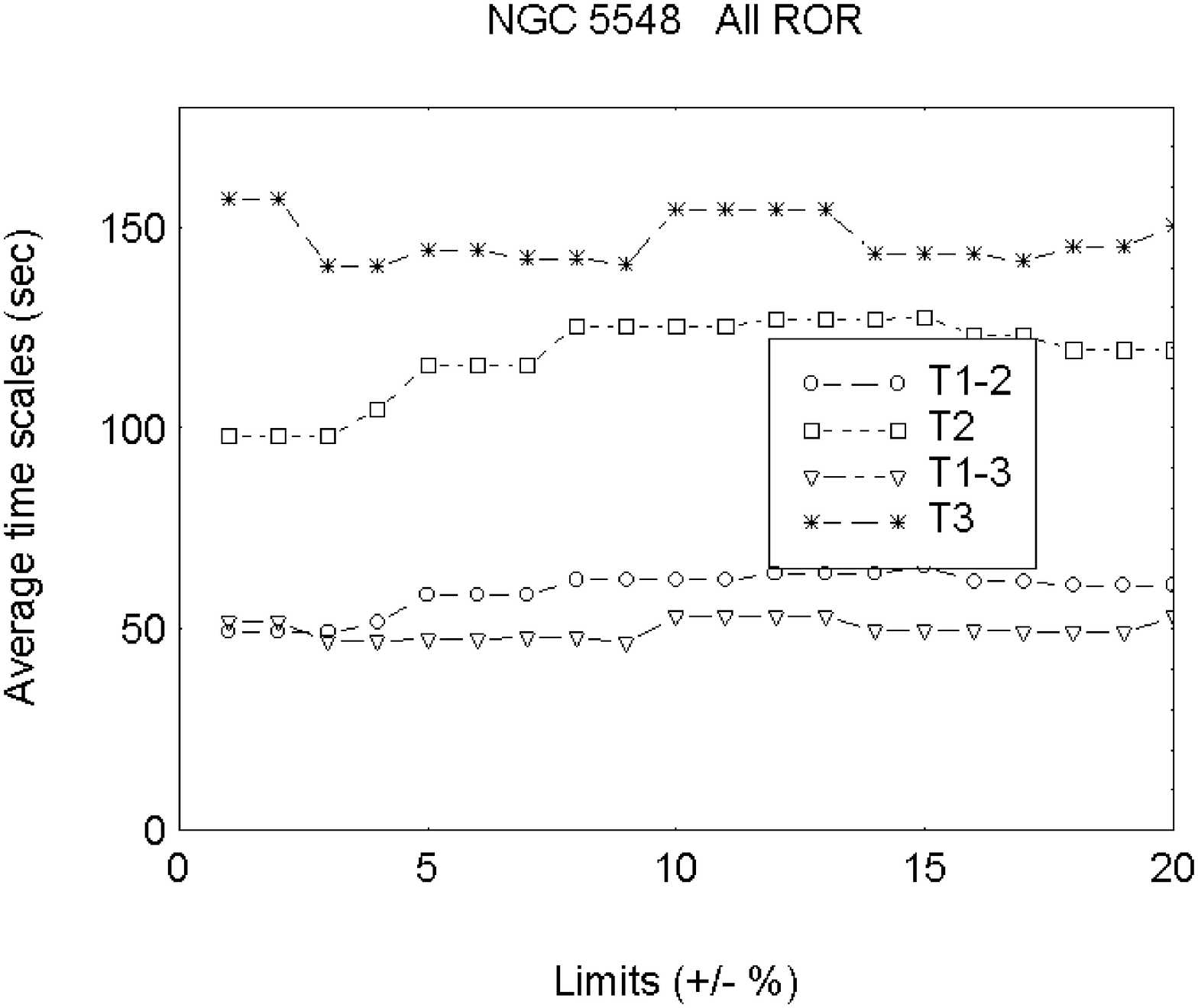}
\caption{Average time scales of peaks identified as multiples as a function of increasing acceptance limits.\label{fig10}}
\end{figure}

{
A single event may be considered as a semi{}-periodic series of pulses. The persistence of a single event is related to its intensity on the time scale spectrum, but may also be studied in a different way. If the series of events is persistent enough it will not only produce a peak at its average time scale, but also at the multiples of the average time scale (sub{}-harmonics). This may be seen in the example of ROR 701246 shown in the left bottom graph of Fig.~\ref{fig6}, where some peaks look, at a first glance, to be the multiples of others. The order of multiples is limited, in addition to the nature of the phenomenon, by the duration of the observation period and subsequently by the largest measurable time scale. The problem may be addressed by a systematic search for multiples (2 and 3 times the basic time scale) when varying the acceptance limit. For a given basic peak at a time scale, T, there is a peak at the multiple time scale, T$_d$, if $|$T$_{d}$ {}- nT)/nT$| < \Delta$, where $\Delta$ is the acceptance limit in percent, a variable in the present analysis, and n = 2, 3 is the number of multiples.
\par}
{
The search for multiples is done for each observation period separately and results are compiled together for all analysed observation periods of NGC 5548. The acceptance limit $\Delta$ was varied from 1 to 20\%, which means that the entire acceptance interval was varied from 2 to 40\%. The result of analysis is given in Fig.~\ref{fig9}, showing number of identified multiples (for both n=2 and n=3) when the value of acceptance limit $\Delta$ increases.
\par}
{
The result of search indicates that, on average, the number of 2T and 3T multiples is nearly the same and that it increases linearly with increasing acceptance interval. Horizontal sections of both curves correspond to empty spaces in the time scale distribution of peaks, where the peaks are obscured by the wobble harmonics. \ It may be shown by simulation that if both 2T and 3T multiples would be equally probable, the 3T curve would be shifted down. The 3T multiples must be more than twice as probable as 2T in order to simulate the result of Fig.~\ref{fig9}.
\par}
{
Another result of simulation is that the wobble harmonics obscure efficiently other spectral peaks deviating less than 4\% of its time scale.
\par}
{
More information about properties of elementary events may be obtained studying average time scales of peaks and their multiples when the acceptance interval increases. The result of averaging is shown in Fig.~\ref{fig10}. T$_{1{}-2 }$is the average time scale of basic peaks, which have been found to correspond to peaks with the double time scale, T$_{2}$. \ T$_{1{}-3}$ is the average time scale of basic peaks, which have been found to correspond to peaks with the triple time scale, T$_{3}$. If the same peaks correspond to both double and triple peaks, T$_{1{}-2}$= T$_{1{}-3}$. If the identification of double peaks is correct, and not a result of too large an acceptance interval, then: T$_{2}$ = 2T$_{1{}-2}$. Similarly: T$_{3}$ = 3T$_{1{}-3}$. It may be seen that all above conditions are fulfilled only up to the acceptance interval of 6\% (acceptance limit of +/{}-3\%). The conclusion must be that the process responsible for generation of elementary events has an internal bandwidth of 6\% (allowed deviations of the time scale).
\par}
{
The length of observation intervals is too short for studying higher order multiples in the spectrum
\par}
\section{A Comparison of Elementary Events in NGC 5548 and in 3C273}
\begin{figure}[p]
\centering
\includegraphics[width=.67\hsize]{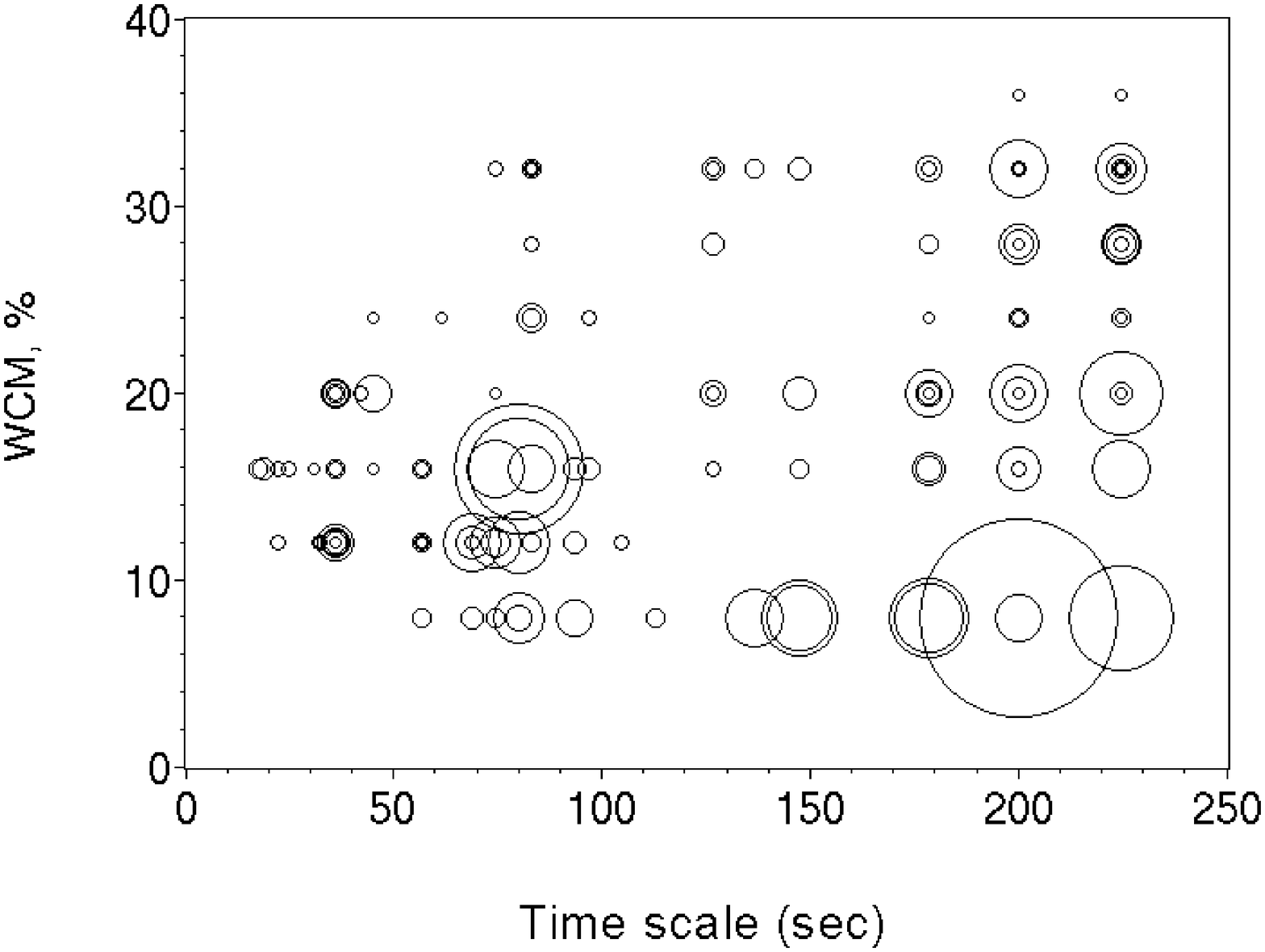}
\caption{The corrected amplitude of elementary events in NGC 5548 as a function of their corrected dominating time scale (in seconds) and the wavelet coefficient magnitude (in \% of the maximum value). The event magnitude is proportional to the diameter of the circle.\label{fig11}}
\end{figure}

\begin{figure}[p]
\centering
\includegraphics[width=.67\hsize,height=.40\hsize]{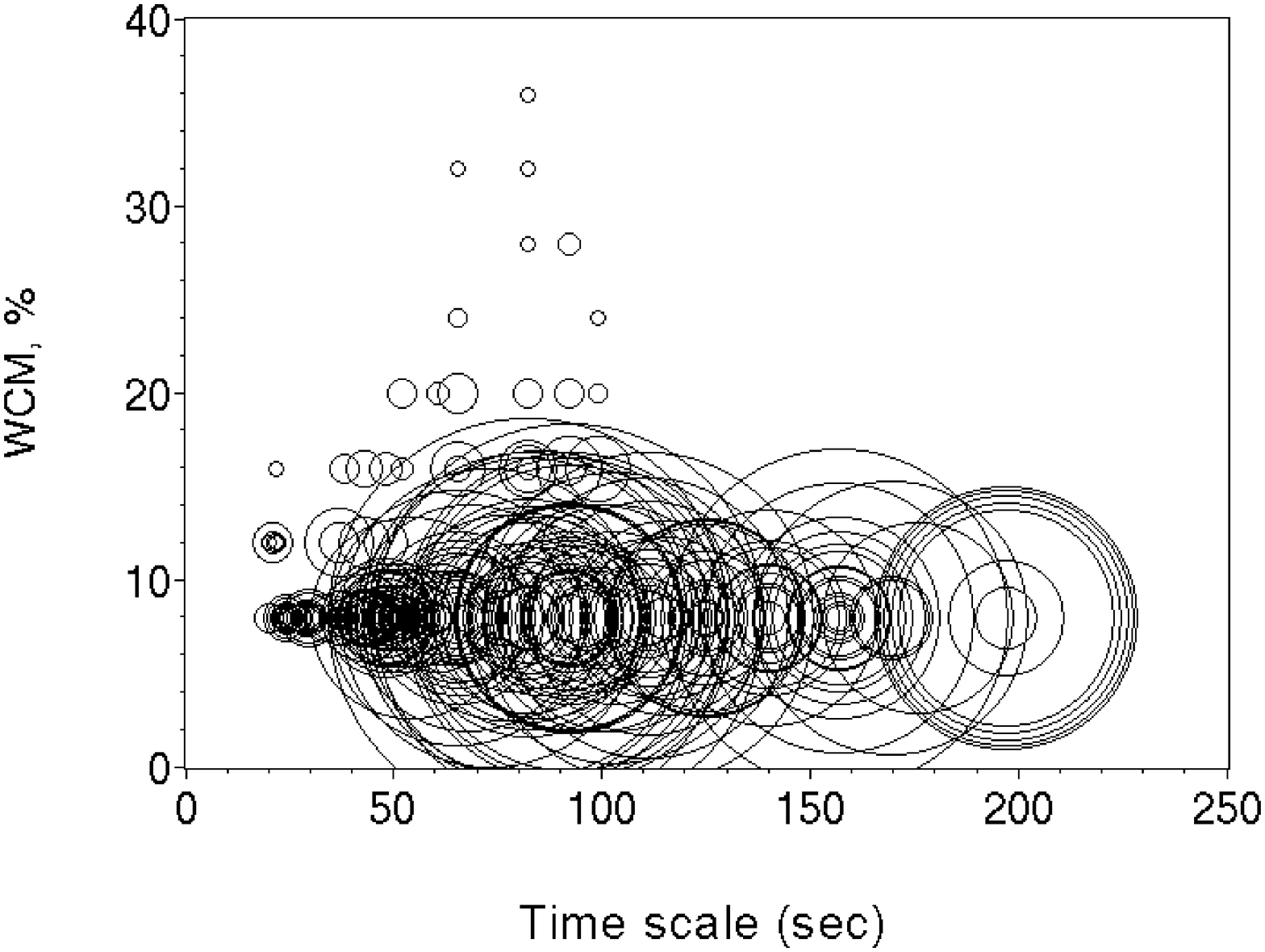}
\caption{The corrected amplitude of elementary events in QSO 3C273 as a function of their corrected dominating time scale (in seconds) and the wavelet coefficient magnitude (in \% of the maximum value). The event magnitude is proportional to the diameter of the circle.\label{fig12}}
\end{figure}

{
It is helpful to compare the occurrence of elementary events in different source types. In order to assure that the observed features are real and not an artefact of the analytical technique, it must be established that the features occur throughout the entire energy range of the PSPC instrument. \ For this purpose, the entire energy range of the PSPC instrument was divided into 2 energy ranges; {\textless}0.5 keV and {\textgreater}0.5 keV. Only events occurring simultaneously at both energy intervals were taken into consideration.
\par}
{
Since there is a substantial difference in the red shift, z, between these two sources, both the time scales and the wavelet coefficient magnitudes must be corrected for the red shift so that the variability spectrum in the source's reference frame will be obtained. \ Both sources are z{}-corrected: the time scale by dividing it with 1+z and the wavelet coefficient magnitudes by multiplying it with (1+z)\textup{\textsuperscript{3/2}}, since the photons are observed within a limited energy interval. Graphs showing the corrected event wavelet coefficient magnitude (diameter of circles) as a function of the corrected time scale (x{}-axis) and of the normalized (to its maximum) wavelet coefficient magnitude (y{}-axis) are shown in Fig.~\ref{fig11} (NGC 5548) and in Fig.~\ref{fig12} (QSO 3C273).
\par}
{
Comparing the above graphs, it is obvious that both the character of the elementary events and their distribution on the time scale {--} wavelet coefficient magnitude plane significantly differs between these two sources. \ Larger amplitudes in the 3C273 quasar mean that the events are more persistent than in the NGC 5548 source. In the 3C273 source, the elementary events are limited to a narrow range of normalized wavelet coefficient magnitudes (around 10\%). This means that elementary events in 3C273 are responsible for not much more than 10\% of its variability. This means that there is more intrinsic Poisson{}-distributed variability of the photon train from that source than from the Seyfert 1 (S1) type source (NGC 5548). As shown earlier, the elementary events are visualized by subtracting the modelled variability from the actual, observed one. As it was shown earlier, the neural network model comprises types of variability common to all data used to construct the model.
\par}
\section{A Comparison of the Modelled Variability in S1 and QSO Sources}
\begin{figure}[p]
\centering
\includegraphics[width=.67\hsize]{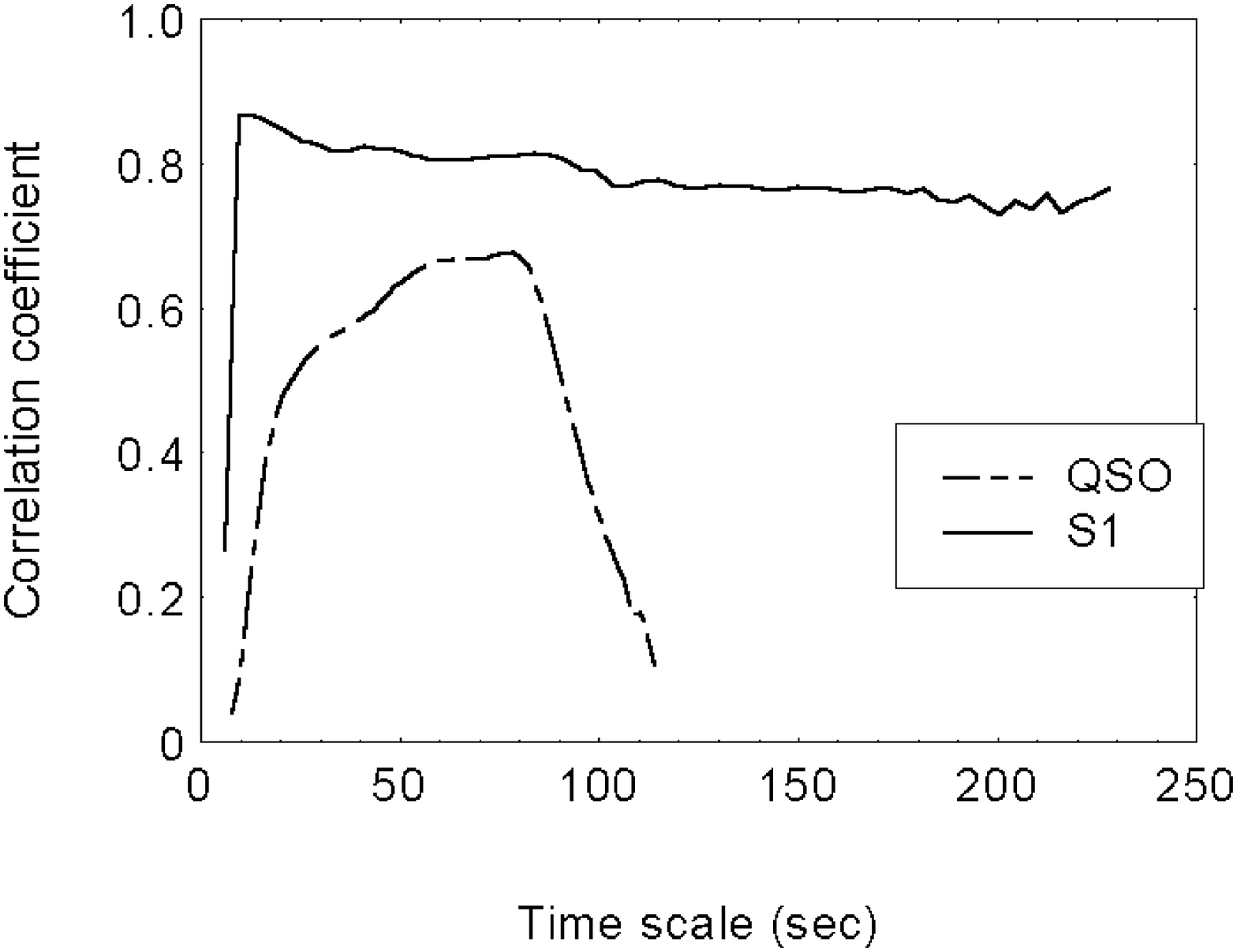}
\caption{Values of correlation coefficients for the 2{}-input network and the observed range of time scales for both source types.\label{fig13}}
\end{figure}

\begin{figure}[p]
\centering
\includegraphics[width=.67\hsize]{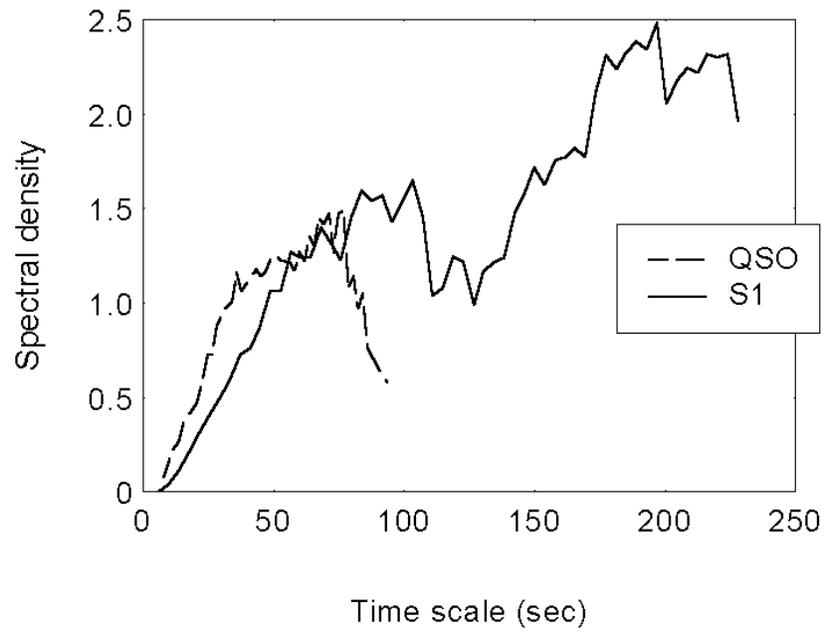}
\caption{The modelled wavelet coefficient magnitude for both source types and the interval of 8 {--} 12\% of the normalized wavelet coefficient magnitude.\label{fig14}}
\end{figure}

{
The above indication of a strong, intrinsic Poisson{}-distributed, variability entails a need for a study comprising a larger number of sources of both types, however excluding sources close to the galactic plane (+/{}-20\textdegree). One model for each source type (27 S1 and 25 QSO) was generated. The models were made only for photon energies {\textless}0.5 keV.
\par}
{
A back{}-propagation (BP) neural network was used for modelling the available data. \ Training was performed using 85\% of the entire data while the remaining 15\% was used to test the model. \ The training data was randomly selected. \ The model consisted of the following variables:
\par}
\paragraph{Input}
\begin{itemize}
\item{Average counts (apparent luminosity) determined for each observation period.\par}
\item{M {--} wavelet coefficient magnitude in \% of its maximum value. 20 levels between 0 and 40\% were used in this study.\par}
\end{itemize}
\paragraph{Output}
{
A 128{}-component spectral vector being that row of the time scale spectrum matrix which corresponds to the actual value of M. Both the time scales and the wavelet filter outputs must be corrected for the red shift, z, so that the variability spectrum in the source's reference frame will be obtained. \ All sources are z{}-corrected: the time scale by dividing with 1+z and the amplitude (wavelet filter output) by multiplying with (1+z)\textsuperscript{3/2}
\par}
{
The BP neural network used to model the data consists of 2 processing elements (PE) in the input layer, 128 PE in the output layer and 150 PE in the hidden layer. The training file contained about 11700 data rows. Recall of the trained network using the test data (not used for training) gives one value of the correlation coefficient for each component of the spectral vector. The correlation coefficient is calculated between the observed vectors and those retrieved from the model. \ This is a measure of both the predictive capability of the model and also of its generalizing ability (response to previously unseen data). \ The values of the correlation coefficients for the present networks (one for each source type) and the observed range of time scales are shown in Fig.~\ref{fig13}.
\par}
{
It may be seen that the modelled variability of both sources may be compared only within a limited interval of time scales: 30 {--} 80 seconds, where both models are reliable enough; the effective observing window.
\par}
{
In order to correct the variability spectrum for the effect of apparent luminosity, the recall of the trained network was done assuming a constant average photon counting rate of 0.1 photons/sec, which is close to the median value for all measured counting rates (0.06 photons/sec). Since the spacecraft wobble is present during all observation periods, it may be expected that the combined effect of the wobble and of the apparent luminosity will be contained in the model together with the intrinsic, persistent variability of the source.
\par}
{
The modelled variability is also strongly dependent on the normalized wavelet coefficient magnitude. Here, recall of the models has been performed for the interval of 8 {--} 12\% of the normalized wavelet coefficient magnitude, where most of the elementary events have been observed. The result of recall is shown in Fig.~\ref{fig14}.
\par}
{
It may be seen that after red shift corrections the modelled variability spectrum within the observing window (30 {--} 80 sec) does not significantly differ between the S1 type and the QSO type sources.
\par}

\section{The Time{}-Domain Analysis of the Short{}-Time Variability}
\begin{figure}[p]
\centering
\includegraphics[width=.67\hsize,height=.27\hsize]{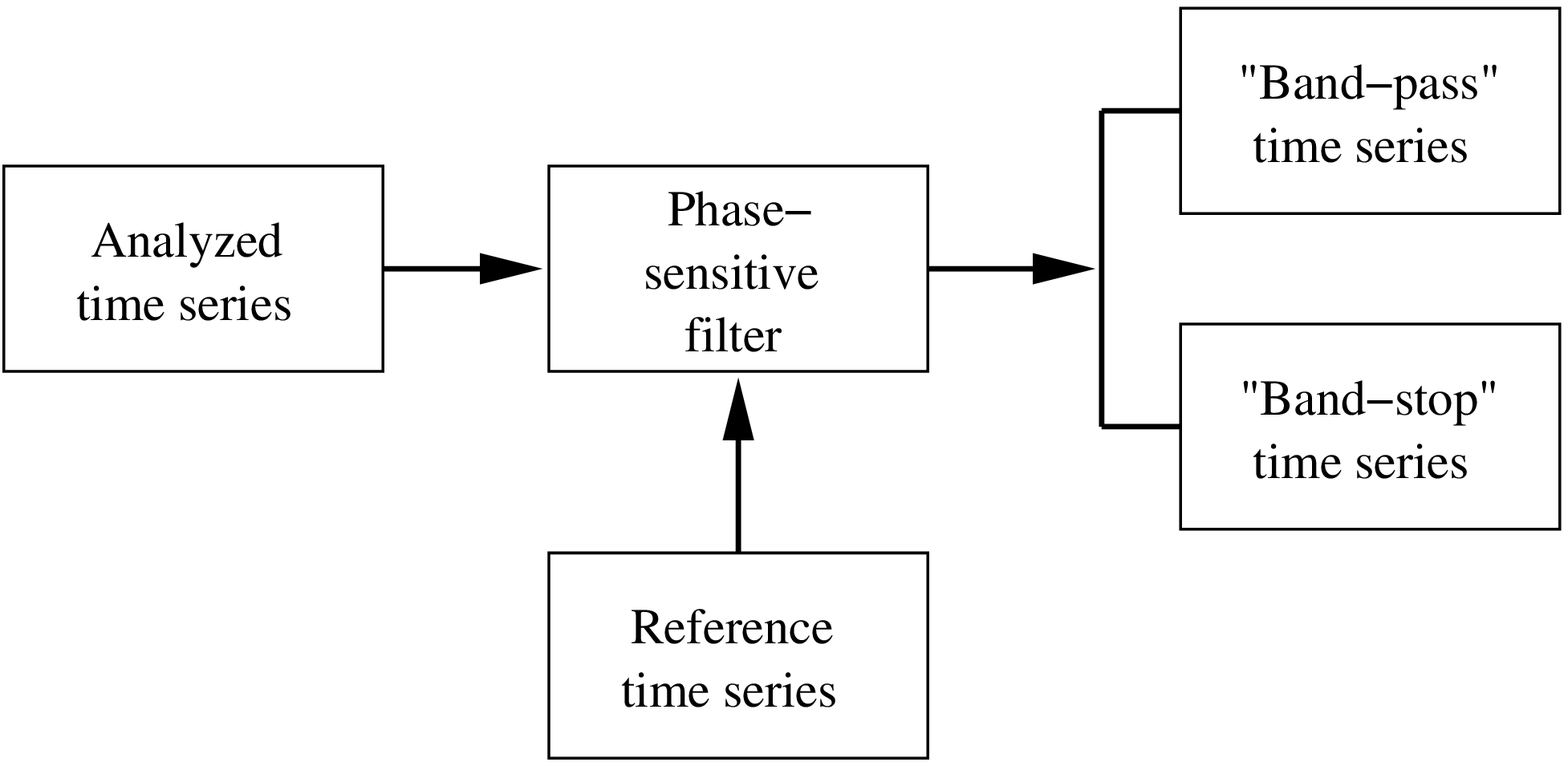}
\caption{The principle of the phase{}-sensitive filter.\label{fig15}}
\end{figure}

\begin{figure}[p]
\centering
\subfigure[The scalogram of the photon counting rate from ROR 701244, NGC 5548]{\includegraphics[width=.47\hsize]{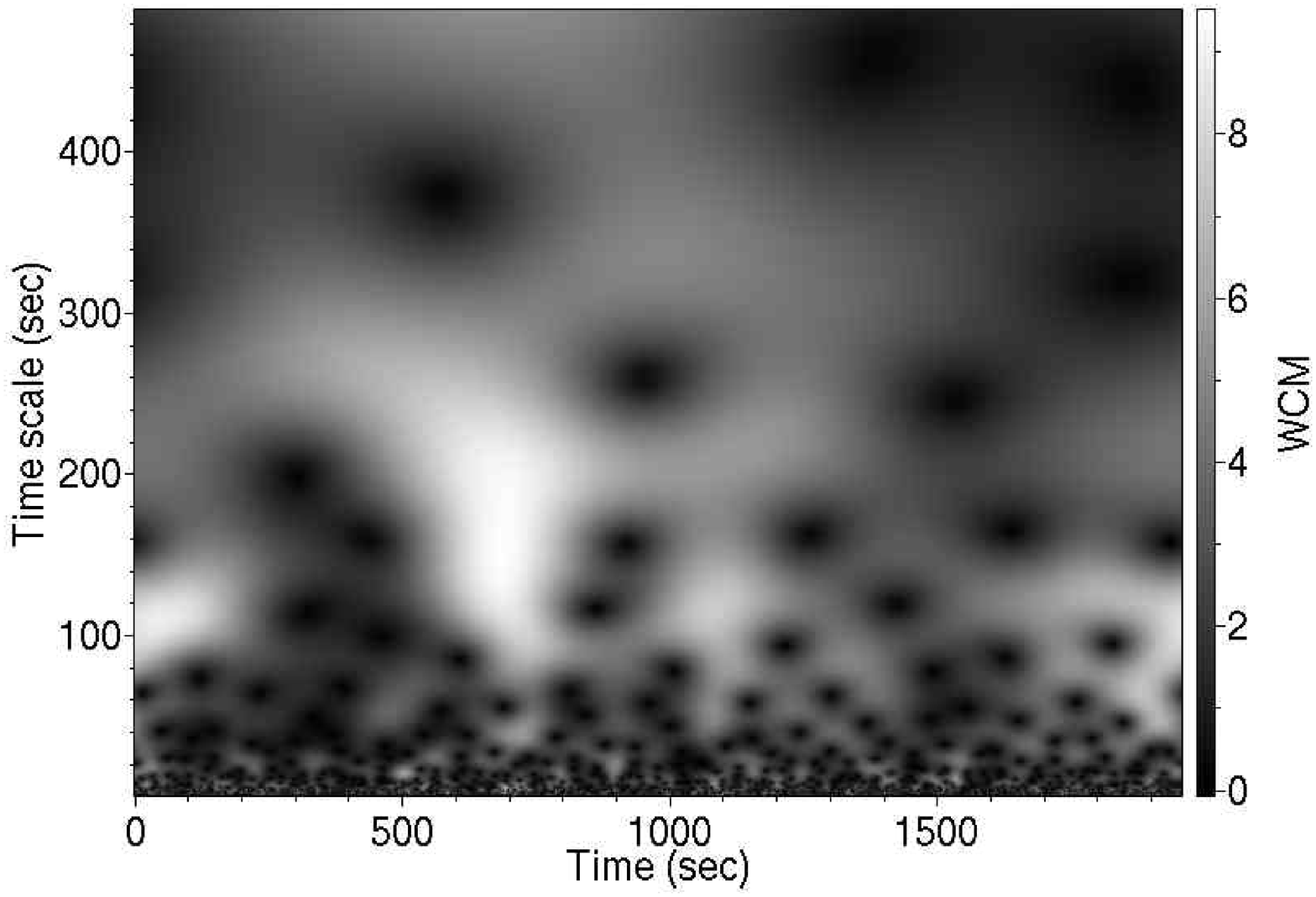}\label{fig16:a}}
\subfigure[The data of Fig.~\ref{fig16:a} from which the wobble{}-related components have been removed. Persistent features of the spectrum were enhanced using the unit{}-variance filter]{\includegraphics[width=.47\hsize]{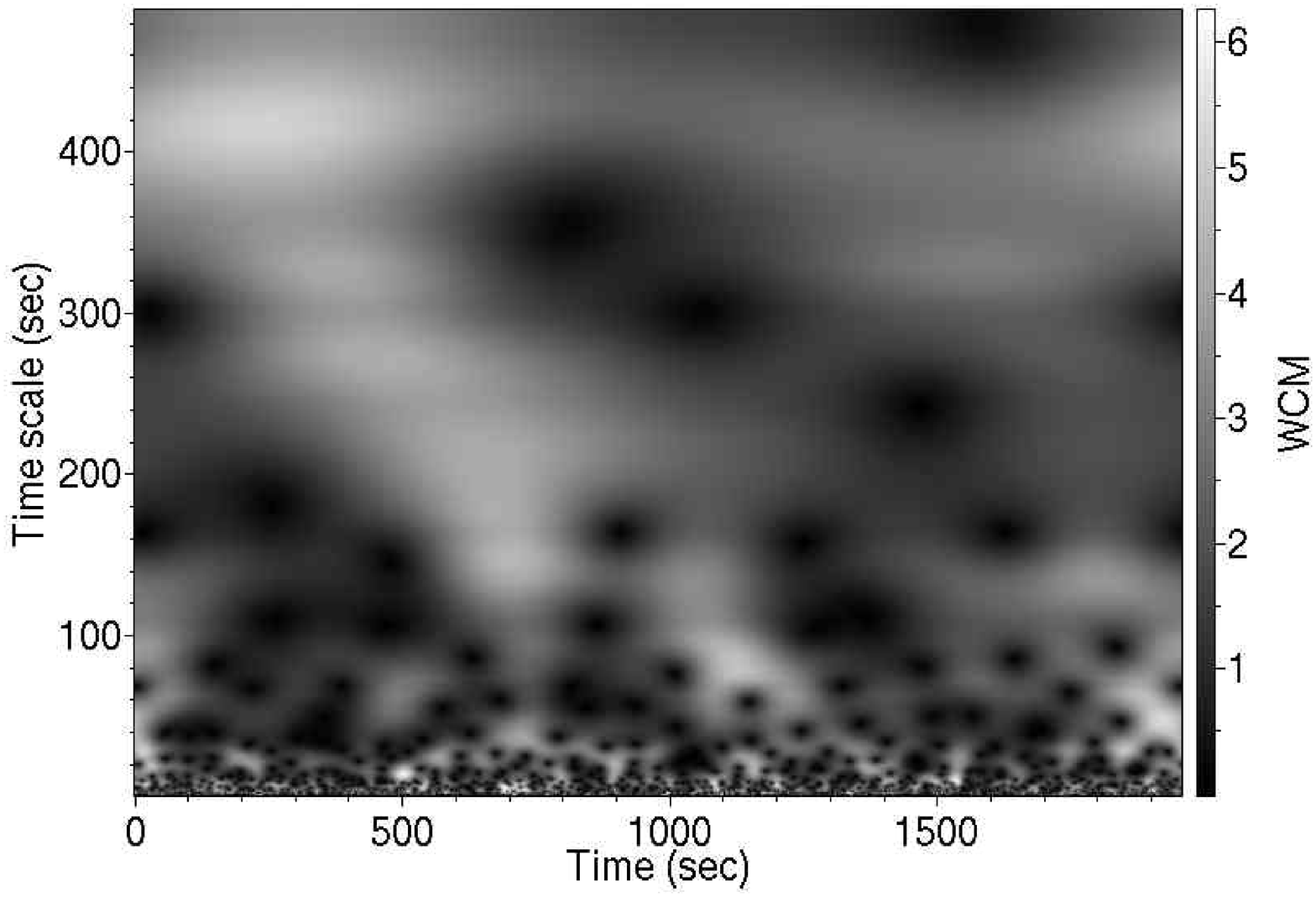}\label{fig16:b}}
\caption{\label{fig16}}
\end{figure}

{
When the investigated signal is polluted by a well known signal, it is possible to remove this unwanted contribution. When a conventional filtering is impossible, the problem may be solved using the wavelet transform. One such application is the influence of the spacecraft wobble on the photon counting rate. Since the photon events are nearly Poisson{}-distributed and the wobble signature contains several harmonics, it is impossible to remove the influence of the wobble using a conventional band{}-stop filter/filters. A possible solution, using the wavelet transform and the accurate information about the wobble phase at different frequencies, was proposed by \citet{liszka03}. The principle of the method is shown on Fig.~\ref{fig15}.
\par}
{
An example showing the original scalogram of photon counting rate from NGC 5548, ROR 701244, is shown in Fig.~\ref{fig16:a}. The scalogram of the same observation file from which the wobble{}-related components have been removed is shown in Fig.~\ref{fig16:b}. The signal has also passed the unit{}-variance filter \citep{liszka03}, which enhances the persistent components in the spectrum.
\par}
\begin{figure}[ptb]
\centering
\subfigure[The scalogram of the photon counting rate from ROR 701256, 3C273]{\includegraphics[width=.47\hsize]{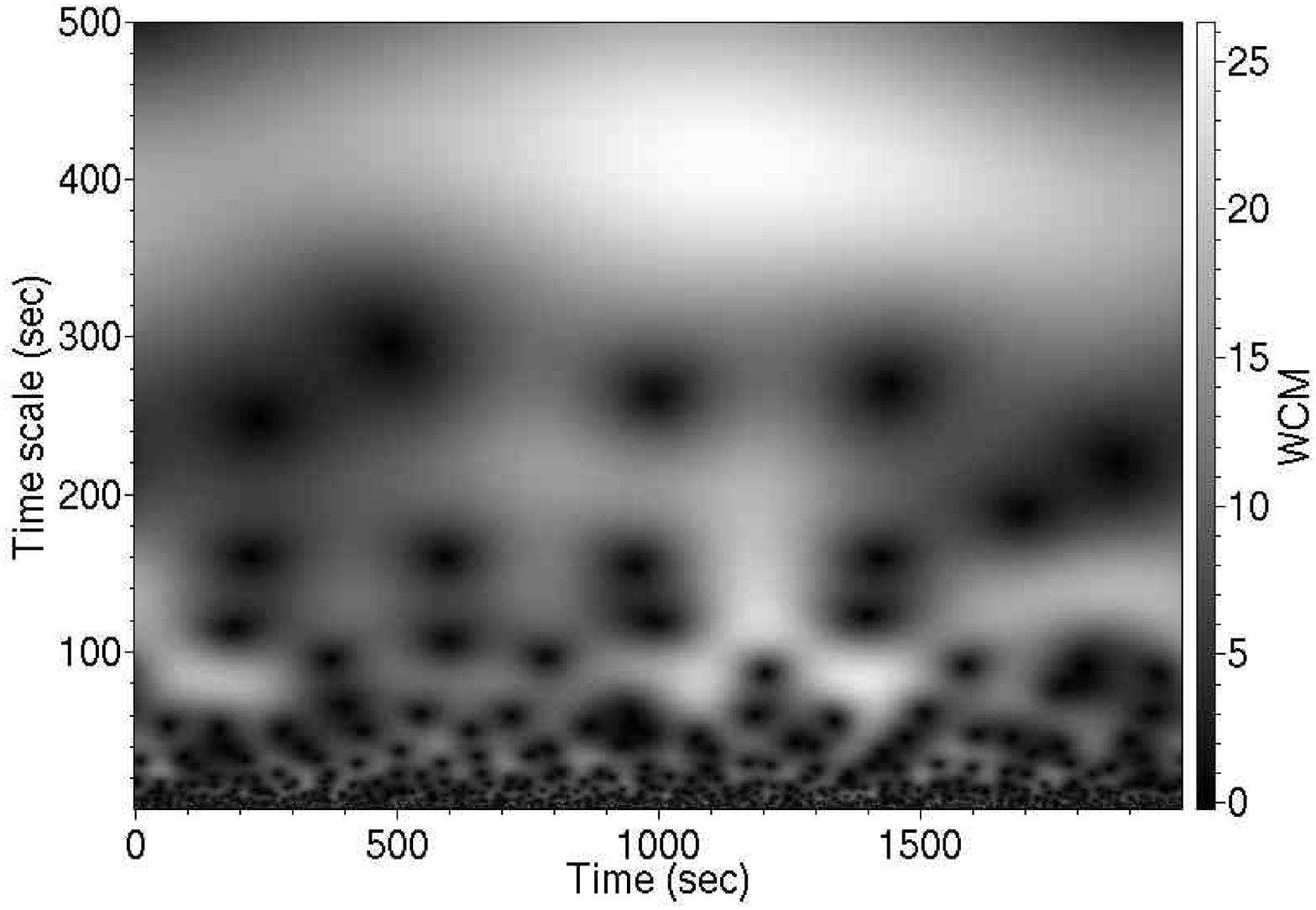}\label{fig17:a}}
\subfigure[The data of Fig.~\ref{fig17:a} from which the wobble{}-related components have been removed. Persistent features of the spectrum were enhanced using the unit{}-variance filter, in picture the pure tone with period of about 300 seconds.]{\includegraphics[width=.47\hsize]{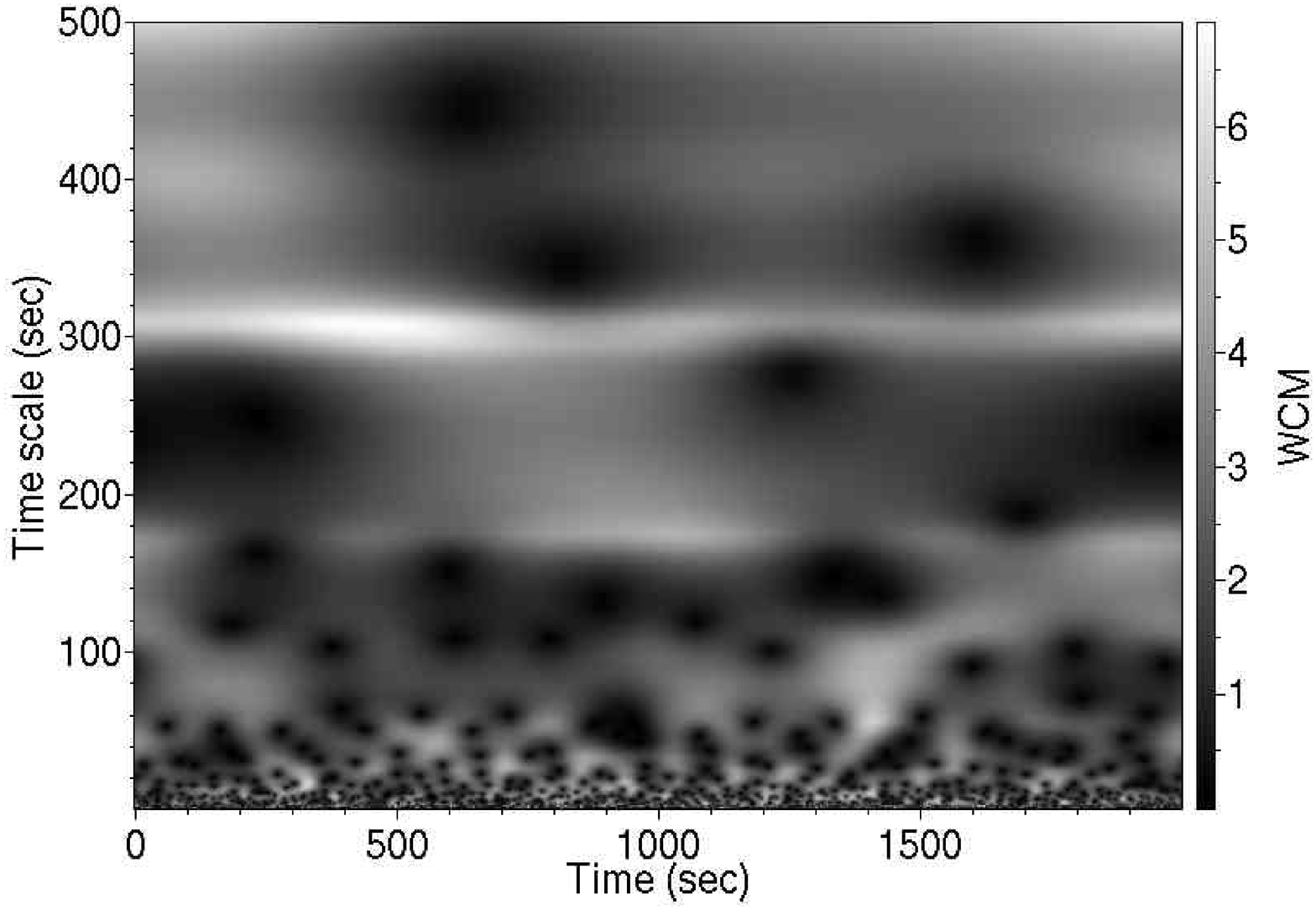}\label{fig17:b}}
\caption{\label{fig17}}
\end{figure}
{
A comparison of images of Figs. \ref{fig16:a} and \ref{fig16:b} shows that the proper filtering of original photon counts reveals new temporal structures in the data. The most prominent structure enhanced is the ``chirp'' (quasi{}-periodicity of increasing frequency) between 200 and 700 seconds of elapsed time.
\par}
{
Also the quasar 3C273 shows numerous deterministic structures after removal of the wobble{}-related components. An example is shown in Figs \ref{fig17:a} and \ref{fig17:b}.
\par}
{
It may be seen that after removal of the wobble{}-related components the 3C273 data reveal deterministic components that can not be related to the wobble. This new technique opens possibilities to study the short{}-time variability of extragalactic sources , even in cases when the instrument is wobbling or ``dithering''.
\par}
\section{Conclusions}
{
We have reviewed and discussed some of the principal obstacles to reducing variability data from astronomical sources {}-{}- including Poisson statistics and instrumental motion, for example satellite wobble {}-{}- and the powerful methods of wavelet transform spectral analysis and non{}-linear filtering for overcoming these. We illustrated their application in a detailed way by applying these techniques to the rapidly variable X{}-ray emission from the Seyfert 1 galaxy NGC 5548 and to the QSO 3C273, confirming the presence and specifying some of the transient quasi{}-periodic and deterministic intrinsic events occurring in these sources.
\par}
{
In particular, after removing the modelled Poisson statistical and wobble{}-related variability components from the ROSAT time{}-scale spectra of 30 observation periods of NGC 5548 X{}-ray data, we found that there is significant residual variability. That is formed by a number of elementary events with a well defined time scale and magnitude. The elementary events seem to be distributed over the entire analyzed range of the time scale spectrum, both in time scale and magnitude domains. There is an indication of increasing intensity towards small magnitudes. Low wobble harmonics suppress, rather than enhance, the observed elementary events. It has been found by simulation that a wobble harmonic efficiently obscure other spectral peaks deviating less than 4\% of its time scale. Typical elementary events last at least 3 basic time scales. The process responsible for the generation of elementary events has an internal bandwidth of 6\% (allowed deviations of the time scale).
\par}
{
For a QSO type source (3C273) the elementary events are confined within a narrow range of normalized wavelet coefficient magnitudes around 10\%. The elementary events in 3C273 are more intense, which may indicate that they are more persistent than the events observed in NGC 5548. The apparent compression of the events around 10\% of the normalized wavelet coefficient magnitude could indicate the existence of a stronger, intrinsic stochastic component in the QSO type of source. However, a study of modelled variability spectra for a larger number of sources indicates that if there is an excess of an intrinsic stochastic component in the QSO sources, it must be less than a factor of 2 (see Fig.~\ref{fig14}).
\par}
{
These methods are widely applicable to all sorts of variable data, most notably to astronomical source variability in other spectral ranges. Certainly, the results presented here need to be further confirmed by applying this analysis to data from more recent X{}-ray satellites, most notably Chandra, and to a larger number of other extragalactic sources.
\par}

%\bibliographystyle{plainnat}
%\bibliography{p10bwi}

\begin{thebibliography}{5}

\bibitem[{Land}(1969)]{land69}
M.~F. {Land}.
\newblock Structure of the retinae of the principal eyes of jumping spiders
  (salticidae: Dendryphantinae) in relation to visual optics.
\newblock {\em Journal of Experimental Biology}, 51:\penalty0 443--470, 1969.

\bibitem[Liszka(2003)]{liszka03}
L.~Liszka.
\newblock {\em Cognitive Information Processing in Space Physics and
  Astrophysics}.
\newblock Pachart Publishing House, 2003.

\bibitem[Liszka and Holmstr{\"o}m(1999)]{liszka99}
L.~Liszka and M.~Holmstr{\"o}m.
\newblock Extraction of a deterministic component from rosat x-ray data using a
  wavelet transform and the principal component analysis.
\newblock {\em \aaps}, 140:\penalty0 125--134, November 1999.

\bibitem[Liszka et~al.(2000a)Liszka, Pacholczyk, and
  Stoeger]{liszka00a}
L.~Liszka, A.G. Pacholczyk, and W.R. Stoeger.
\newblock Extraction of a deterministic component from rosat x-ray data
  using a wavelet transform and the principal component analysis. ii. the data
  analysis.
\newblock {\em \aap}, 354:\penalty0 847--852, February 2000a.

\bibitem[Liszka et~al.(2000b)Liszka, Pacholczyk, and
  Stoeger]{liszka00b}
L.~Liszka, A.G. Pacholczyk, and W.R. Stoeger.
\newblock Active galactic nuclei. vi. rosat variability of seyfert galaxies.
\newblock {\em \apj}, 540:\penalty0 122--130, September 2000b.

\end{thebibliography}
\end{document}